\documentclass{aa}

\usepackage{graphicx}
\usepackage{txfonts}
\usepackage[allcolors=blue]{hyperref}
\makeatletter
\renewcommand*\aa@pageof{, page \thepage{} of \pageref*{LastPage}}
\makeatother

\def \rsun{R$\mathrm{_\odot}$}
\def \msun{M$\mathrm{_\odot}$}

\def \teff{$T_\mathrm{eff}$\,}
\def \logg{$\log g$}
\def \MH{$\left[M/H\right]$}
\def \prot{$P_\mathrm{rot}$}
\def \vsini{$v\,\mathrm{sin}\,i$}
\def \MG{M$_{\mathrm{G}}$}
\def \BP{G$_{\mathrm{BP}}$}
\def \RP{G$_{\mathrm{RP}}$}
\def \kms{km\,s$^{-1}$\,}

\def \1s{$1\,\sigma$}
\def \kid{$\chi^2$}
\def \t0{T$_0$}

\def \vmic{$\upsilon_{\mathrm mic}$\,}
\def \geff{$g_{\rm eff}$\,}

\newcommand{\orcidlink}[1]{\protect\href{https://orcid.org/#1}{\protect\includegraphics[width=8pt]{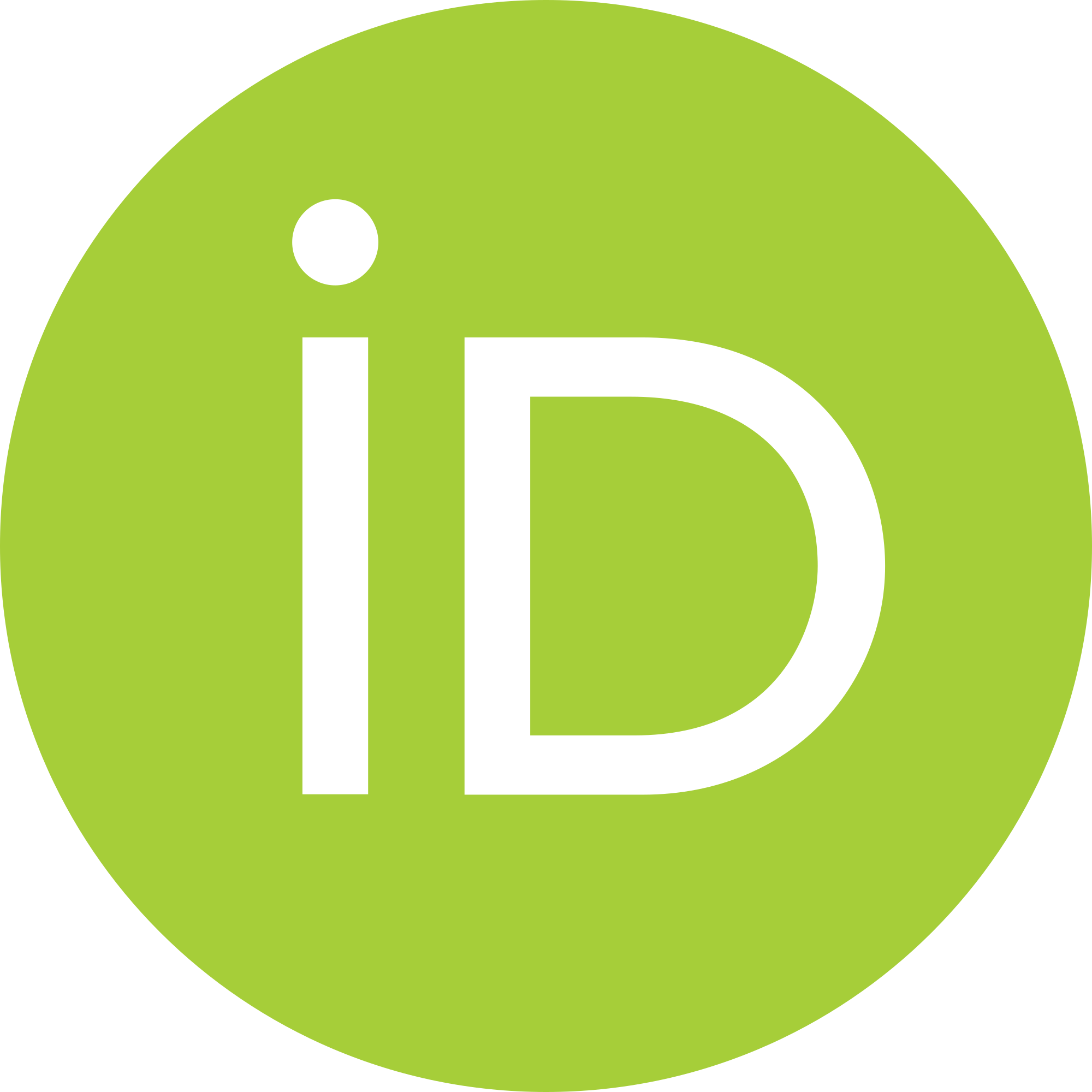}}}

\begin{document}

   \title{The SPIRou legacy survey}

   \subtitle{Rotation period of quiet M dwarfs from circular polarization in near-infrared spectral lines: I. The SPIRou APERO analysis}

   \author{P. Fouqu\'{e}\inst{1}\orcidlink{0000-0002-1436-7351}
          \fnmsep\thanks{Based on observations obtained at the Canada-France-Hawaii Telescope (CFHT) which is operated from the summit of Maunakea by the National Research Council of Canada, the Institut National des Sciences de l'Univers of the Centre National de la Recherche Scientifique of France, and the University of Hawaii. Based on observations obtained with SPIRou, an international project led by Institut de Recherche en Astrophysique et Plan\'etologie, Toulouse, France.}
          \and
          E. Martioli\inst{2,3}\orcidlink{0000-0002-5084-168X}
          \and
          J.-F. Donati\inst{1}\orcidlink{0000-0001-5541-2887}
          \and
          L.T. Lehmann\inst{1}\orcidlink{0000-0001-7264-0734}
          \and
          B. Zaire\inst{1,4}\orcidlink{0000-0002-9328-9530}
          \and
          S. Bellotti\inst{1,5}\orcidlink{0000-0002-2558-6920}
          \and
          E. Gaidos\inst{6}\orcidlink{0000-0002-5258-6846}
          \and
          J. Morin\inst{7}
          \and
          C. Moutou\inst{1}\orcidlink{0000-0002-2842-3924}
          \and
          P. Petit\inst{1}\orcidlink{0000-0001-7624-9222}
          \and
          S.H.P. Alencar\inst{4}
          \and
          L. Arnold\inst{8}\orcidlink{0000-0002-0111-1234}
          \and
          \'E. Artigau\inst{9}\orcidlink{0000-0003-3506-5667}
          \and
          T.-Q. Cang\inst{10}
          \and
          A. Carmona\inst{11}\orcidlink{0000-0003-2471-1299}
          \and
          N.J. Cook\inst{9}\orcidlink{0000-0003-4166-4121}
          \and
          P. Cort\'es-Zuleta\inst{12}\orcidlink{0000-0002-6174-4666}
          \and
          P.I. Cristofari\inst{1}
          \and
          X. Delfosse\inst{11}\orcidlink{0000-0001-5099-7978}
          \and
          R. Doyon\inst{9}\orcidlink{0000-0001-5485-4675}
          \and
          G. H\'ebrard\inst{3}\orcidlink{0000-0001-5450-7067}
          \and
          L. Malo \inst{9}
          \and
          C. Reyl\'e\inst{13}\orcidlink{0000-0003-2258-2403}
          \and
          C. Usher\inst{8}
          }

   \institute{
        \inst{1} Institut de Recherche en Astrophysique et Plan\'{e}tologie, Universit\'{e} de Toulouse, CNRS, 14 avenue Edouard Belin, F-31400, Toulouse, France,
        \email{Pascal.Fouque@irap.omp.eu}\\
        \inst{2} Laborat\'{o}rio Nacional de Astrof\'{i}sica, Rua Estados Unidos 154, 37504-364, Itajub\'{a} - MG, Brazil\\
        \inst{3} Institut d'Astrophysique de Paris, CNRS, Sorbonne Universit\'{e}, 98 bis bd Arago, 75014 Paris, France\\
        \inst{4} Departamento de F\'{i}sica-Icex-UFMG Ant\^{o}nio Carlos, 6627, 31270-901 Belo Horizonte, MG, Brazil \\
        \inst{5} Science Division, Directorate of Science, European Space Research and Technology Centre (ESA/ESTEC), Keplerlaan 1, 2201 AZ, Noordwijk, The Netherlands\\
        \inst{6} Department of Earth Sciences, University of Hawai’i at M\={a}noa, 1680 East-West Road, Honolulu, HI, 96822, USA\\
        \inst{7} Laboratoire Univers et Particules de Montpellier, Universit\'e de Montpellier, CNRS, F-34095 Montpellier, France\\
        \inst{8} Canada-France-Hawaii Telescope, CNRS, Kamuela, HI 96743, USA\\
        \inst{9} Universit\'e de Montr\'eal, D\'epartement de Physique, IREX, Montr\'eal, QC H3C 3J7, Canada\\
        \inst{10} Beijing Normal University, No.19, Xinjiekouwai St, Haidian District, Beijing, 100875, People's Republic of China\\
        \inst{11} Universit\'e Grenoble Alpes, CNRS, IPAG, F-38000 Grenoble, France\\
        \inst{12} Laboratoire d'Astrophysique de Marseille, Aix-Marseille Universit\'e, CNRS, 38 rue Fr\'ed\'eric Joliot-Curie, F-13388, Marseille, France\\
        \inst{13} Institut UTINAM, CNRS, Universit\'e Bourgogne Franche-Comt\'e, OSU THETA Franche-Comt\'e-Bourgogne, Observatoire de Besan\c{c}on, BP 1615, 25010 Besan\c{c}on Cedex, France
        }
   \date{Received ; accepted }

 
  \abstract
   {The rotation period of stars is an important parameter together with mass, radius, and effective temperature. It is an essential parameter for any radial velocity monitoring, as stellar activity can mimic the presence of a planet at the stellar rotation period. Several methods exist to measure it, including long sequences of photometric measurements or temporal series of stellar activity indicators.}
   {Here, we use the circular polarization in near-infrared spectral lines for a sample of 43 quiet M dwarfs and compare the measured rotation periods to those obtained with other methods.}
   {From Stokes $V$ spectropolarimetric sequences observed with SPIRou at the Canada-France-Hawaii Telescope and the data processed with the APERO pipeline, we computed the least-squares deconvolution profiles using different masks of atomic stellar lines with known Land\'e factor appropriate to the effective temperature of the star. We derived the longitudinal magnetic field to examine its possible variation in 50 to 200 observations of each star. To determine the stellar rotation period, we applied a Gaussian process regression, enabling us to determine the rotation period of stars with evolving longitudinal field.}
   {We were able to measure a rotation period for 27 of the 43 stars of our sample. The rotation period was previously unknown for 8 of these stars. Our rotation periods agree well with periods found in the literature based on photometry and activity indicators, and we confirm that near-infrared spectropolarimetry is an important tool for measuring rotation periods, even for magnetically quiet stars. Furthermore, we computed the ages for 20 stars of our sample using gyrochronology.}
   {}

   \keywords{stars: planetary systems --  stars: individual -- techniques: polarimetry, radial velocity}

   \maketitle
%

\section{Introduction}

The rotation period (hereafter \prot) is an important characteristic of a star, as is its effective temperature (\teff), mass, radius and corresponding gravity (\logg), metallicity (\MH), and age. In the context of exoplanet search by velocimetry (measure of the radial velocity, RV), the knowledge of the stellar rotation period avoids attributing a periodogram peak to an exoplanet orbital period by taking the systematic noise that is introduced by stellar activity into account \citep[e.g.,][]{queloz01}. Several publications identified a periodic signal as due to activity and/or proposed methods to distinguish RV jitter from Keplerian variation (see, e.g., \citealt{bonfils07, figueira10, boisse11}) or identified a signal that was previously announced as due to a planet as actually being caused by activity (e.g., \citealt{huelamo08, robertson14a, robertson14b, kane16, faria20}).

Stellar rotation periods for magnetic stars have been measured using the small-scale surface magnetic field, which is detected through the broadening and intensification of some stellar lines due to the Zeeman effect \citep[e.g.,][]{babcock47} or the large-scale surface magnetic field derived from the polarimetric signal of spectral lines, as pioneered by \citet{preston71, landstreet80} for early-type stars. A periodic variation in the longitudinal magnetic field is interpreted as due to the periodic appearance of magnetic regions at the surface of the star, which allows measuring its rotation period; see \citet{landstreet92} for a review. An application to M dwarfs can be found in \citet{donati08, morin08, morin10, hebrard16}.

Rotation periods can also be measured from long photometric sequences obtained with small telescopes such as ASAS \footnote{All Sky Automated Survey}\citep{pojmanski97}, APT \footnote{Automatic Photoelectric Telescope}\citep{henry99}, HATNet \footnote{Hungarian-made Automated Telescope Network}\citep{bakos04}, NSVS \footnote{Northern Sky Variability Survey}\citep{wozniak04}, SuperWASP \footnote{Wide Angle Search for Planets}\citep{pollacco06}, and MEarth \citep{NC08, irwin09}: examples can be found in \citet{ks07, irwin11, suarez16, diez19}, among others, or more recently, from space telescopes such as Kepler, K2, or TESS \footnote{Transiting Exoplanet Survey Satellite}\citep{diez19}.

An alternative technique uses sequences of stellar activity indicators such as H$\alpha$, Ca II H\&K, the Ca II infrared triplet, or values computed at the same time as the radial velocities using the cross-correlation function (CCF) such as the full width at half maximum (FWHM), contrast, or the bisector \citep[e.g.,][]{noyes84, queloz01, bonfils07, west08, boisse11, bonfils13, suarez15, nelson16, suarez18, toledo19, lafarga21}.

In the framework of the SPIRou legacy survey (hereafter SLS), a sample of about 50 nearby low-mass stars is regularly monitored to detect exoplanets using the SPIRou spectropolarimeter \citep{donati20}. Some of these targets lack a measured rotation period, and some have several measurements that  sometimes disagree. To clarify the situation and increase the sample of known rotation periods, we focus in this paper on determining the rotation period by exploiting the large-scale magnetic field information extracted from the Stokes $V$ profiles of spectral lines.

The paper is organized as follows: Section \ref{sec:observations} briefly describes SPIRou\footnote{ \url{http://spirou.irap.omp.eu} and \url{https://www.cfht.hawaii.edu/Instruments/SPIRou/}} and the SPIRou Legacy Survey (SLS\footnote{\url{http://spirou.irap.omp.eu/Observations/The-SPIRou-Legacy-Survey}}; id P40 and P42, PI: Jean-Fran\c{c}ois Donati). Section \ref{sec:reductionandanalysis} describes the generic software called {\it A PipelinE to Reduce Observations} APERO (\footnote{\url{https://github.com/njcuk9999/apero-drs}} v0.7.232, \citealt{cook22}), which is used to reduce the SPIRou observations, in particular, the spectropolarimetric analysis and the derivation of the longitudinal magnetic field. Section \ref{sec:results} explains the results for the stars in our sample, separating them into three categories according to their spectral type. Two examples are treated in detail, and a comparison is made with results from the literature. Finally, Section \ref{sec:conclusions} examines our general results and further perspectives.

\section{Observations}
\label{sec:observations}

Observations are part of the CFHT large program SPIRou Legacy Survey. SPIRou is a stabilized high-resolution near-infrared (NIR) spectropolarimeter \citep{donati20} mounted on the 3.6~m CFHT at the top of Maunakea, Hawaii. It is designed for high-precision velocimetry and spectropolarimetry to detect and characterize exoplanets and stellar magnetic fields. It provides full coverage of the NIR spectrum from 950~nm to 2500~nm at a spectral resolving power of $\lambda / \Delta \lambda \sim 70\,000 $. 

The SLS was allocated 310 nights over seven semesters (February 2019 to June 2022). It covers three different science topics, called work packages (see \citealt{donati20} for details). The analysis of this paper is restricted to work package 1 (WP1), dedicated to a blind planet search. More than 50 M dwarfs were regularly monitored during the SPIRou runs (typically ten contiguous nights per month). We excluded monitored active stars with short and well-known rotation periods from this study (\prot < 5~d): Gl\,388 (AD\,Leo), Gl\,406 (CN\,Leo), Gl\,873 (EV\,Lac), GJ\,1111 (DX\,Cnc), GJ\,1154, GJ\,1245B, GJ\,3622, and PM\,J18482+0741. We therefore selected 43 stars in the original WP1 sample for which about 150 (50 at least, 250 at most) polarimetric sequences (of four individual subexposures each) have been secured. This represents about 6800 visits, corresponding to more than 27\,000 individual spectra secured for the considered sample. The stars belonging to our sample and their stellar characteristics are given in Table \ref{tab:samplechars}: {\it Gaia} absolute magnitudes and colors use {\it Gaia} DR3 \citet{gaia21}; effective temperatures and metallicities come from \citet{cristofari22}, who used the same set of observations and the same sample (except for Gl\,581, which does not satisfy our criterion of the minimum number of visits); and masses were computed from absolute 2MASS $K_{\rm s}$ magnitude and metallicity using the \citealt{mann19} relations. The relative precision of these masses is 2-3\% according to \citet{mann19}, and the internal errors amount to 30\,K and 0.1\,dex for \teff and [M/H], according to \citet{cristofari22}. We divided our sample into three subsamples according to a proxy of stellar mass. These three mass bins correspond to the three types of magnetic behavior identified among active M dwarfs by \citet{morin10} (see Figure~15 therein). We prefer to define regions according to the absolute $G$ magnitude rather than using the spectral types. Although they match on average, a low metallicity makes the star fainter at a given spectral type: Gl\,412A (M1.0V, \MH$=-0.42$) is 1\,mag fainter in  $G$ than Gl\,410 (M1V, \MH$=+0.05$), therefore both should not belong to the same group. The horizontal lines in Table \ref{tab:samplechars} separate the three subsamples. The number of visits corresponds to the initial number, before some polarimetric sequences were rejected for the reasons explained in Sec. \ref{sec:spectropolarimetry}. Finally, we list the FWHM of the median Stokes $I$ profile of each star because it was used to define the velocity range on which we measured the longitudinal magnetic field from the Stokes $V$ profile, as explained in Sec. \ref{sec:spectropolarimetry}.

   \begin{table*}
    \caption[]{Stellar characteristics of the sample of 43 M dwarfs. For uncertainties on mass, \teff , and [M/H], see text. The three sections correspond to early-, mid-, and late-type M dwarfs from top to bottom}
    \label{tab:samplechars}
    \begin{center}
    \begin{tabular}{lcccccccc}
        \hline
        \noalign{\smallskip}
 Star & spectral type & $M_{\rm G}$ & $G_{\rm BP} - G_{\rm RP}$ & mass & $T_{\rm eff}$ & [M/H] & visits & FWHM Stokes $I$ \\
 Units & & mag & mag & \msun & K & dex & & \kms \\
        \noalign{\smallskip}
        \hline
        \noalign{\smallskip}
Gl\,338B & M0V & 8.046 & 1.846 & 0.58 & 3952 & $-0.08$ & 58 & $6.95 \pm 0.25$ \\
Gl\,846 & M0.5V & 8.282 & 1.967 & 0.57 & 3833 & 0.07 & 194 & $6.65 \pm 0.26$ \\
Gl\,205 & M1.5V & 8.327 & 2.122 & 0.58 & 3771 & 0.43 & 160 & $6.54 \pm 0.16$ \\
Gl\,410 & M1V & 8.426 & 2.006 & 0.55 & 3842 & 0.05 & 131 & $7.73 \pm 0.21$ \\
Gl\,880 & M1.5V & 8.611 & 2.151 & 0.55 & 3702 & 0.26 & 166 & $6.37 \pm 0.17$ \\
Gl\,514 & M1.0V & 8.793 & 2.088 & 0.50 & 3699 & $-0.07$ & 177 & $5.84 \pm 0.29$ \\
Gl\,382 & M2.0V & 8.898 & 2.234 & 0.51 & 3644 & 0.15 & 124 & $6.34 \pm 0.31$ \\
        \noalign{\smallskip}
        \hline
        \noalign{\smallskip}
Gl\,752A & M3.0V & 9.240 & 2.379 & 0.47 & 3558 & 0.11 & 130 & $6.06 \pm 0.23$ \\
Gl\,48 & M2.5V & 9.364 & 2.459 & 0.46 & 3529 & 0.08 & 194 & $5.99 \pm 0.26$ \\
Gl\,617B & M3V & 9.459 & 2.483 & 0.45 & 3525 & 0.20 & 149 & $6.09 \pm 0.21$ \\
Gl\,412A & M1.0V & 9.460 & 2.104 & 0.39 & 3620 & $-0.42$ & 224 & $5.50 \pm 0.24$ \\
Gl\,15A & M1.5V & 9.460 & 2.164 & 0.39 & 3611 & $-0.33$ & 256 & $5.76 \pm 0.26$ \\
Gl\,849 & M3V & 9.511 & 2.542 & 0.46 & 3502 & 0.35 & 203 & $6.17 \pm 0.22$ \\
Gl\,411 & M2.0V & 9.522 & 2.216 & 0.39 & 3589 & $-0.38$ & 182 & $4.93 \pm 0.18$ \\
Gl\,480 & M3.5V & 9.565 & 2.591 & 0.45 & 3509 & 0.26 & 104 & $6.14 \pm 0.22$ \\
Gl\,436 & M3.0V & 9.631 & 2.449 & 0.42 & 3508 & 0.03 & 90 & $5.89 \pm 0.33$ \\
Gl\,687 & M3.0V & 9.739 & 2.518 & 0.39 & 3475 & 0.01 & 227 & $5.69 \pm 0.20$ \\
Gl\,408 & M2.5V & 9.831 & 2.430 & 0.38 & 3487 & $-0.09$ & 179 & $6.17 \pm 0.33$ \\
Gl\,317 & M3.5V & 9.859 & 2.664 & 0.42 & 3421 & 0.23 & 79 & $5.39 \pm 0.21$ \\
GJ\,4063 & M3.5V & 9.982 & 2.770 & & 3419 & 0.42 & 220 & $5.96 \pm 0.22$ \\
Gl\,725A & M3.0V & 10.119 & 2.461 & 0.33 & 3470 & $-0.26$ & 219 & $5.43 \pm 0.22$ \\
Gl\,251 & M3.0V & 10.129 & 2.561 & 0.35 & 3420 & $-0.01$ & 187 & $5.73 \pm 0.26$ \\
GJ\,4333 & M3.5V & 10.233 & 2.818 & 0.37 & 3362 & 0.25 & 193 & $5.70 \pm 0.16$ \\
GJ\,1012 & M4V & 10.268 & 2.710 & 0.35 & 3363 & 0.07 & 142 & $5.26 \pm 0.17$ \\
GJ\,1148 & M4V & 10.370 & 2.778 & 0.34 & 3354 & 0.11 & 105 & $5.45 \pm 0.16$ \\
Gl\,876 & M3.5V & 10.527 & 2.809 & 0.33 & 3366 & 0.15 & 88 & $5.65 \pm 0.28$ \\
PM J09553-2715 & M3V & 10.629 & 2.667 & 0.29 & 3366 & $-0.03$ & 75 & $5.77 \pm 0.25$ \\
PM J08402+3127 & M3.5V & 10.739 & 2.703 & 0.28 & 3347 & $-0.08$ & 142 & $5.33 \pm 0.25$ \\
Gl\,725B & M3.5V & 10.790 & 2.625 & 0.25 & 3379 & $-0.28$ & 212 & $5.34 \pm 0.22$ \\
GJ\,1105 & M4V & 10.931 & 2.792 & 0.27 & 3324 & $-0.04$ & 171 & $5.44 \pm 0.17$ \\
Gl\,445 & M3.5V & 10.948 & 2.702 & 0.24 & 3356 & $-0.24$ & 94 & $5.21 \pm 0.19$ \\
GJ\,3378 & M3.5V & 10.975 & 2.791 & 0.26 & 3326 & $-0.05$ & 177 & $5.65 \pm 0.22$ \\
Gl\,169.1A & M4V & 10.994 & 2.896 & 0.28 & 3307 & 0.13 & 185 & $5.66 \pm 0.28$ \\
        \noalign{\smallskip}
        \hline
        \noalign{\smallskip}
GJ\,1289 & M3.5V & 11.556 & 3.011 & 0.21 & 3238 & 0.05 & 209 & $7.60 \pm 0.37$ \\
PM J21463+3813 & M4V & 11.591 & 2.814 & 0.18 & 3305 & $-0.38$ & 187 & $5.21 \pm 0.26$ \\
GJ\,1103 & M4.5V & 11.817 & 3.116 & 0.19 & 3170 & $-0.03$ & 70 & $5.44 \pm 0.28$ \\
Gl\,699 & M4.0V & 11.884 & 2.834 & 0.16 & 3311 & $-0.37$ & 249 & $6.04 \pm 0.32$ \\
Gl\,15B & M3.5V & 11.928 & 2.836 & 0.16 & 3272 & $-0.42$ & 189 & $6.33 \pm 0.37$ \\
Gl\,447 & M4.0V & 11.960 & 3.033 & 0.18 & 3198 & $-0.13$ & 60 & $6.24 \pm 0.24$ \\
GJ\,1151 & M4.5V & 12.158 & 3.142 & 0.17 & 3178 & $-0.16$ & 157 & $6.42 \pm 0.37$ \\
Gl\,905 & M5.5V & 12.881 & 3.529 & 0.15 & 3069 & 0.05 & 220 & $6.13 \pm 0.29$ \\
GJ\,1286 & M5.5V & 13.344 & 3.706 & 0.12 & 2961 & $-0.23$ & 113 & $7.46 \pm 0.46$ \\
GJ\,1002 & M5.5V & 13.347 & 3.675 & 0.12 & 2980 & $-0.33$ & 146 & $6.92 \pm 0.34$ \\
        \noalign{\smallskip}
        \hline
    \end{tabular}
    \end{center}
  \end{table*}

\section{SPIRou data reduction and analysis} 
\label{sec:reductionandanalysis}

\subsection{ APERO reduction}
\label{sec:aperoreduction}

Our SPIRou data were reduced with the software APERO. APERO first performs some initial preprocessing of the 4096$\times$4096~pixel images of the HAWAII\,4RG$^{\rm {\small TM}}$ (H4RG), applying a series of procedures to correct detector effects, remove background thermal noise, and identify bad pixels and cosmic-ray impacts. 

It then uses exposures of a quartz halogen lamp (flat) to calculate the position of 49 of the 50 \'echelle spectral orders recorded on the detector. It optimally  extracts \citep{horne86} spectra of the two science channels (fibers A and B) and the simultaneous reference channel (fiber C). This APERO extraction takes the non-Gaussian shape of the instrument profile generated by the pupil slicer into account. APERO corrects the spectra for the blaze signature of the \'echelle orders obtained from the flat-field exposures, as described in \citet{cook22}. Both a 2D order-by-order and 1 D order-merged spectrum are produced for each channel of each scientific exposure.

The pixel-to-wavelength calibration is obtained from exposures of both a uranium-neon hollow cathode lamp and a Fabry-P\'erot etalon, generally following the procedure given in \citet{hobson21}, but with differences described in \citet{cook22}. We refer to Sections 5.4 and 6.6 of the latter paper for details about the use of a reference night for the whole survey and updates for each survey night, with subtle differences in the wavelength calibration of fibers AB with respect to fibers A and B separately. This procedure provides wavelengths in the rest frame of the observatory, but APERO also calculates the barycentric Earth radial velocity (BERV) and the barycentric Julian date (BJD) of each exposure using the code \texttt{barycorrpy}\footnote{\url{https://github.com/shbhuk/barycorrpy}}\citep{wright14, kanodia18}. These can then be used to reference the wavelength and time to the barycentric frame of the Solar System. 

APERO calculates the spectrum of the telluric transmission using a novel technique based on a model obtained from the collection of standard star observations carried out since the beginning of SPIRou operations in 2019 and a fit  made for each individual observation using a revised technique based on the principles explained in \cite{artigau14}, as briefly described in \citet{cook22} and in more details in \citet{artigau22}. APERO also calculates the Stokes parameters describing the polarization state, as defined, for instance, in \citet{landi92}. Each orthogonal polarization state is recorded on one science fiber. Four sub-exposures are taken with different positions of the quarter-wave rhombs: the method is described in \cite {donati97}. The two polarization states of each subexposure are combined to define the Stokes parameters, as explained in Sec. 10.1 of \citet{cook22}) using equations defined in \citet{bagnulo09}: the Stokes $I$ parameter measures the intensity,  the Stokes $V$ parameter measures the circular polarization, and the Stokes $Q$ and $U$ measure the linear polarization (not used in thiswork). Another combination defines a null measurement to quantify the quality of the Stokes $V$ detection. The continuum for Stokes $I$ and $V$ can be modeled by an iterative sigma-clip algorithm to fit a polynomial to the data. The spectra are normalized to this continuum before an LSD analysis is performed. More details are given in \citet{martioli20, cook22}.

\subsection{Spectropolarimetry analysis}
\label{sec:spectropolarimetry}

We further analyzed the SPIRou polarized spectra using the code \texttt{spirou-polarimetry}\footnote{\url{https://github.com/edermartioli/spirou-polarimetry}} , which we applied to the telluric-corrected 2D spectra. We assumed that the APERO telluric correction is good enough so that we did not need to exclude zones of telluric water bands in this code. The Stokes $I$, Stokes $V$, and null-polarization spectra were compressed to one-line profiles using the least-squares deconvolution (LSD) method of \cite{donati97}. The line masks used in our LSD analysis were computed using the VALD3 catalog \citep{piskunov95, ryabchikova15} and a MARCS model atmosphere \citep{gustafsson08} with a grid of effective temperatures between 3000~K and 4000~K by steps of 500~K, a constant surface gravity of $\log g=5.0$~dex, and a micro-turbulent velocity \vmic of 1 \kms. We selected all lines deeper than 3\% and with a Land\'{e} factor of $g_{\rm eff}>0$ for a total of 2460, 1335, and 956 atomic lines for 4000\,K, 3500\,K, and 3000~K, respectively. We normalized the Stokes $V$ profile using a mean wavelength, a mean depth, and a mean Land\'e factor of the lines in the mask, but we only used the mean depth to normalize the Stokes $I$ profile. For the 4000\,K mask (3500~K and 3000~K), the mean values are 1651.2\,nm (1617.9\,nm and 1604.8\,nm) for the wavelengths, 0.135 (0.133 and 0.155) for the depths, and 1.249 (1.242 and 1.235) for the Land\'e factors. To ensure that Stokes $I$ LSD profiles have a continuum at 1, we used a linear regression of a significant number of bins located well outside of the line for each profile to improve the continuum normalization. We measured the standard deviation of the values in these bins and rejected the profiles for which this standard deviation significantly deviated from the average over all the profiles. This linear regression gives us a better defined Voigt profile, allowing a more precise measure of the FWHM of the median Stokes $I$ LSD profile of each star.

Not all polarimetric sequences led to a useful measurement of the longitudinal magnetic field. There are several steps of data rejection: the telluric correction of some spectra may fail, some LSD profiles are noisy enough that they cannot be fit by a Voigt profile, and others can be fit, but present a noisier continuum than the rest (see above) and are thus rejected. We only rejected data that might contaminate our results, especially, to build a clean median Stokes $I$ profile on which we measured the FWHM and defined the velocity limits to integrate the Stokes $V$ profile (using 6 FWHM as the measuring window, meaning $\pm 3$ FWHM, generally about $\pm 20$ \kms). The final number of polarimetric sequences used in our study is about 6500, compared to the 6800 initially available sequences for our sample of stars. This is an acceptable loss of about 4\% that generally corresponds to spectra with a low signal-to-noise ratio (S/N).

\begin{figure*}
   \centering
   \includegraphics[width=\hsize]{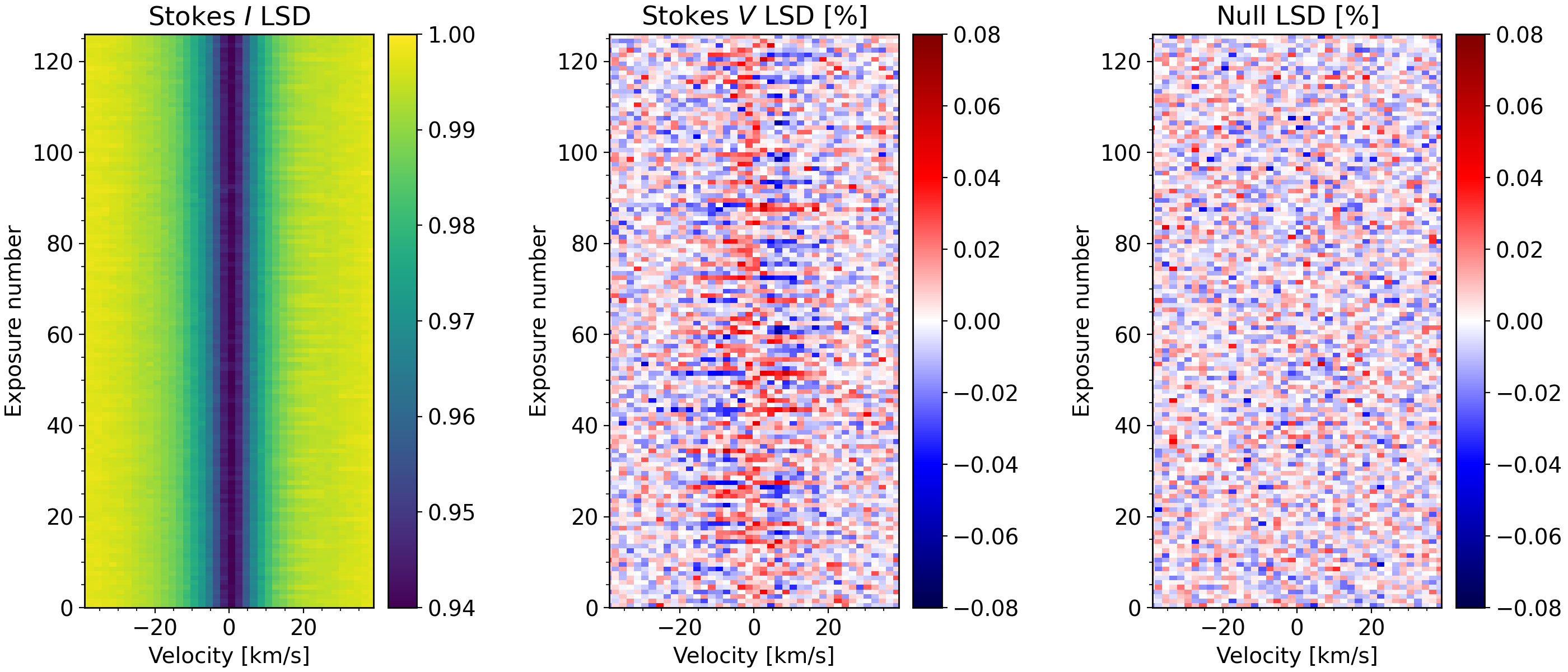}
      \caption{Dynamical maps of Gl\,410 constructed from individual observations (horizontal bands), stacked vertically. The first plot shows Stokes $I$ LSD profiles, where the continuum is given in yellow and the different absorption depths are shown in green and blue shades. The following plots show the Stokes $V$ and null LSD profiles, where positive values are given as red shades, and negative values are shown as blue shades. Note that Stokes $V$ and null LSD amplitudes are expressed in percent. All plots are given in the rest frame of Gl\,410 and display a velocity window of 10 FWHM (see Table~\ref{tab:samplechars}).
       }
        \label{fig:GL410_lsd_timeseries}
  \end{figure*}

\begin{figure}
   \centering
   \includegraphics[width=\hsize]{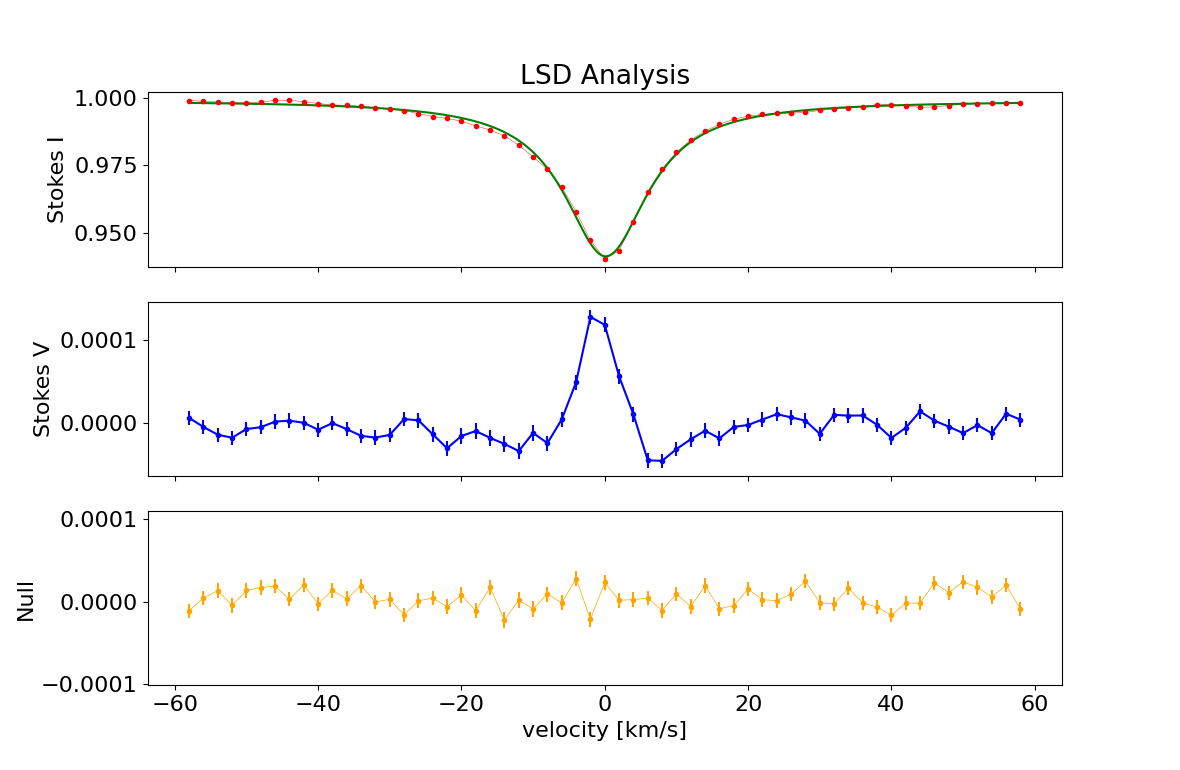}
      \caption{Median of all LSD profiles in the Gl\,410 SPIRou time series. The top panel shows Stokes $I$ LSD (red points) with a Voigt profile model fit (green line), the middle panel shows Stokes $V$ (blue points), and the bottom panel shows the null polarization profile (orange points). 
       }
        \label{fig:GL410_lsd_median_profiles}
  \end{figure}

\begin{figure*}
   \centering
   \includegraphics[width=\hsize]{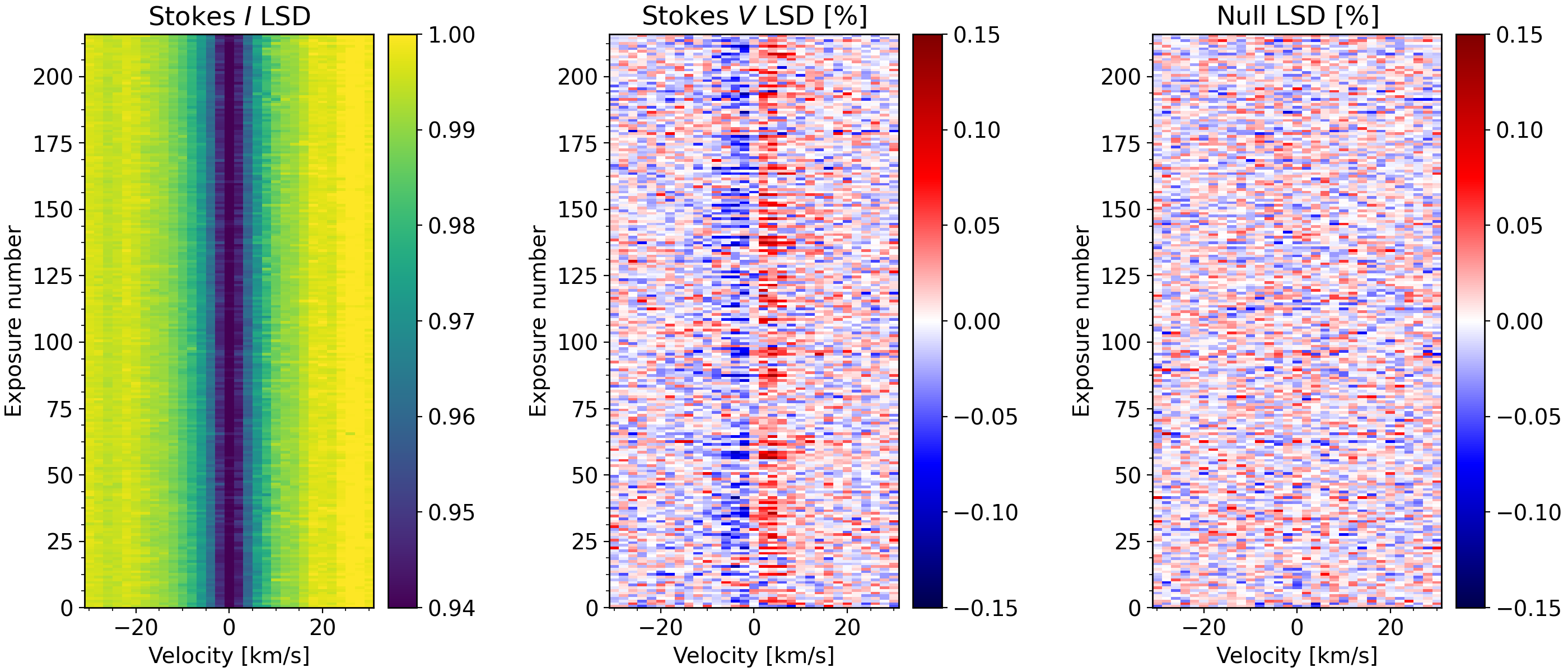}
      \caption{Similar to Fig.~\ref{fig:GL410_lsd_timeseries}, but for the dynamical maps of Stokes $I$, $V$ and null LSD profiles collected for Gl\,905 using SPIRou.
       }
        \label{fig:GL905_lsd_timeseries}
  \end{figure*}

\begin{figure}
   \centering
   \includegraphics[width=\hsize]{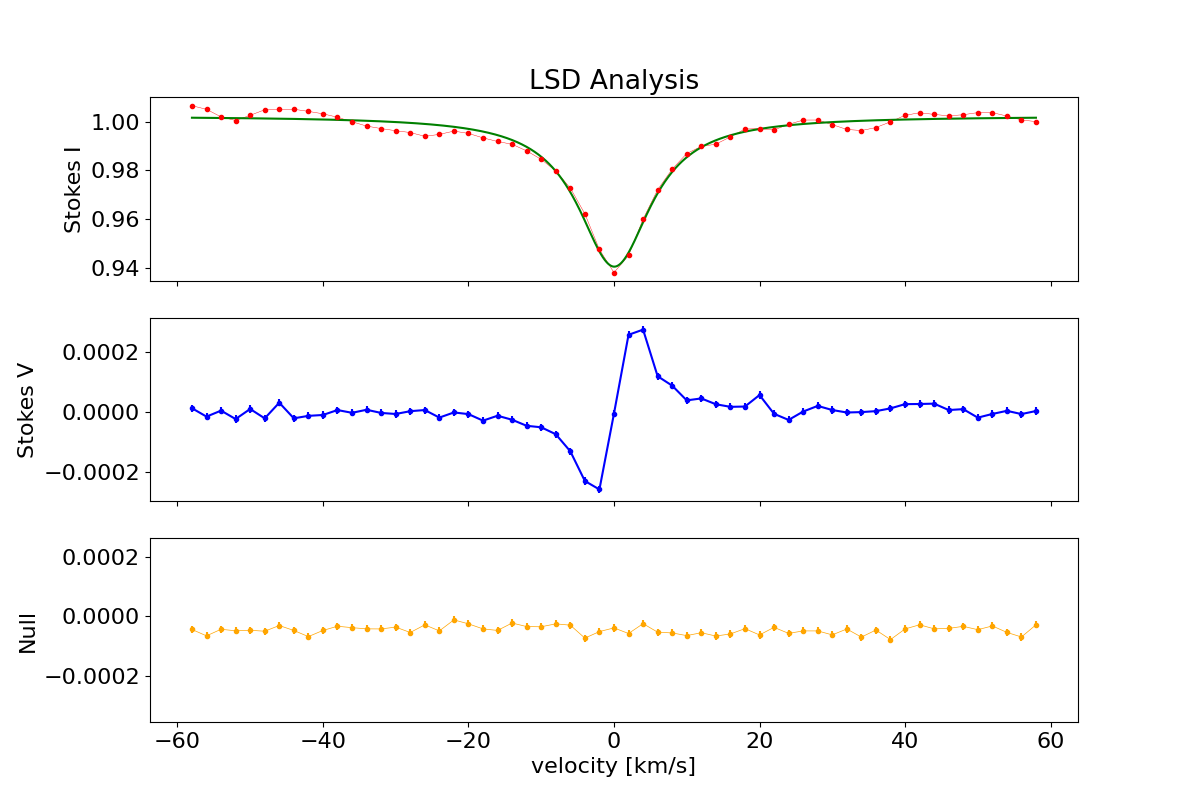}
      \caption{Median of all LSD profiles in the Gl\,905 SPIRou time series. The top panel shows Stokes $I$ LSD (red points) with a Voigt profile model fit (green line), the middle panel shows Stokes $V$ (blue points), and the bottom panel shows the null polarization profile (orange points). 
       }
        \label{fig:GL905_lsd_median_profiles}
  \end{figure}

As an example, we took the LSD profiles of the early-M dwarf Gl\,410 (DS\,Leo), which displays mild signs of magnetic activity. Fig.~\ref{fig:GL410_lsd_timeseries} shows the resulting LSD profiles at each observing epoch, stacked vertically. Dynamical maps from left to right correspond to the intensity (Stokes $I$), the circular polarization (Stokes $V$), and the null LSD profiles. A nonzero Stokes $V$ profile indicates the presence of magnetic fields, while the null profile allows us to assess the reliability of the Stokes $V$ detection. Finally, Fig. \ref{fig:GL410_lsd_median_profiles} shows the median profile of our time series, where the Zeeman signature is clearly visible in the detection of the Stokes $V$ profile. We also display a median null profile to assess that the Stokes $V$ signature is reliably detected. For comparison, we display the Stokes $I$, $V$ and null dynamical maps (Fig.~\ref{fig:GL905_lsd_timeseries}) and median profiles (Fig.~\ref{fig:GL905_lsd_median_profiles}) of the late-M dwarf Gl\,905, whose longitudinal magnetic field has a similar amplitude as Gl\,410. Here again, the Stokes $V$ signal is clearly detected, with a shape corresponding at first order to the first derivative of the Stokes $I$ profile.

In the appendix, we display the Stokes $V$ temporal evolution for each star of our sample. The Stokes $I$ temporal evolution and the median Stokes profiles carry less important information for the present work, and the null profiles are very similar to each other. We decided not to publish them for brevity.

To confirm the consistency of our measurements, we also obtained an independent polarimetric reduction and LSD analysis of our SPIRou data using the pipeline \texttt{Libre-Esprit}  \citep{donati97,donati20}. The results of this alternative analysis will be given in a forthcoming paper (Donati et al., in prep).

\subsection{Longitudinal magnetic fields}
\label{sec:stellaractivity}

To diagnose the magnetic field in our sample of stars, we calculated the longitudinal magnetic field $B_\ell$ in the LSD profiles of SPIRou following the same prescription as in \cite{rees79, donati97, moutou20, martioli20}. $B_\ell$ is defined as the brightness-weighted line-of-sight-projected component of the vector magnetic field integrated over the visible hemisphere of the star, and is given in G by

\begin{equation}
    \label{eq:Bl}
    B_\ell=-2.142 \times 10^{11} \frac{\int v \, V (v) \, dv}{\lambda_0 \cdot g_{\rm eff} \cdot c \cdot \int \left[ 1 - I(v) \right] dv},
\end{equation}
where $c$ is the speed of light in the same unit as $v$ (\kms), $I(v)$ is the Stokes $I$ LSD profile normalized by the continuum, and $V(v)$ is the Stokes $V$ LSD profile, both as functions of the velocity $v$ in the stellar frame, $\lambda_{0}$ is the mean wavelength in nm, and $g_{\rm eff}$ is the mean Land\'e factor of the lines included in the LSD analysis.

\begin{figure}
   \centering
       \includegraphics[width=1.0\hsize]{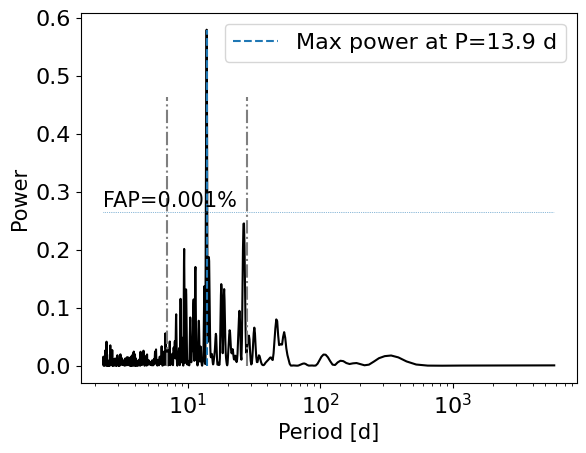}
      \caption{Generalized Lomb-Scargle periodogram analysis of the longitudinal magnetic field ($B_\ell$) time series of Gl\,410. The dashed blue line shows the highest power at a period of 13.9~d, and the dot-dashed lines indicate possible harmonics at half and twice the period.
      }
        \label{fig:GL410_blong_periodogram}
  \end{figure}

\begin{figure}
   \centering
       \includegraphics[width=1.0\hsize]{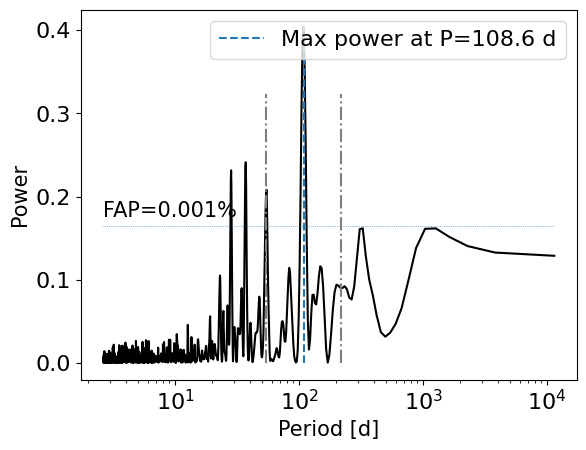}
      \caption{GLS periodogram analysis of the longitudinal magnetic field ($B_\ell$) time series of Gl\,905. The dashed blue line shows the highest power at a period of 108.6~d, and the dot-dashed lines indicate possible harmonics at half and twice the period.
      }
        \label{fig:GL905_blong_periodogram}
  \end{figure}

As the spatial distribution of the magnetic regions may be nonaxisymmetric, $B_\ell$ can be modulated by the stellar rotation. This allows deriving the stellar rotation period if any periodicity is detected in its time series \citep[e.g.,][]{preston71, landstreet80}. Examples focusing more on K and M dwarfs can be found in \citep[e.g.,][]{morin08, moutou17, petit21}.  Fig. \ref{fig:GL410_blong_periodogram} and Fig. \ref{fig:GL905_blong_periodogram} show the generalized Lomb-Scargle (GLS) periodogram \citep{zechmeister09} for the $B_\ell$ data calculated using the tool \texttt{astropy.timeseries}\footnote{\url{https://docs.astropy.org/en/stable/timeseries/lombscargle.html}}  for our two example stars Gl\,410 and Gl\,905. We find a maximum power at a period of 13.9~d for Gl\,410 and at 108.6~d for Gl\,905, both with a false-alarm probability (FAP) below 0.001\%. In the case of Gl\,905, three other peaks below an FAP of 0.001\% can be noted in the periodogram, and they are due to harmonics of the main period.

As the magnetic field of M dwarfs is likely to evolve with time, this study requires a flexible model to account for the variability. We employed a Gaussian process (GP) regression analysis \citep[e.g.,][]{haywood14,aigrain15} using the code \texttt{george}\footnote{\url{http://dfm.io/george/current/}} \citep{ambikasaran15}, where we assumed that the rotationally modulated stellar activity signal in $B_\ell$ is quasi-periodic (QP). Thus, we adopted a parameterized covariance function (or kernel) as in \cite{angus18}, which is given by

\begin{equation}
    k(\tau_{ij}) = \alpha^{2} \exp{\left[ -\frac{\tau_{ij}^2}{2l^2} - \frac{1}{\beta^2} \sin^{2}{\left( \frac{\pi \tau_{ij}}{P_{\rm rot}} \right)} \right]} + \sigma^{2} \delta_{ij},
\end{equation}

\noindent where $\tau_{ij} = t_{i} - t_{j}$ is the time difference between data points $i$ and $j$, $\alpha^{2}$ is the amplitude of the covariance, $l$ is the decay time, $\beta$ is the smoothing factor, $P_{\rm rot}$ is the star rotation period, and $\sigma$ is an additional uncorrelated white noise, which adds a jitter term to the diagonal of the covariance matrix.  This kernel combines a squared exponential component describing the overall covariance decay and a component that describes the periodic covariance structure, the amplitude of which is controlled by the smoothing factor\footnote{The definition of the decay timescale and smoothing factor vary among researchers. The former is sometimes defined as $\sqrt{2} \, l$ \citep[e.g.,][]{petit21}, and $1/\beta^2$ may appear as $\Gamma$ and be called harmonic complexity \citep[e.g.,][]{nicholson22}.}. Typical values of $\beta$ vary between 0.25 and 1.25. The lower values correspond to multiple harmonics, and the higher values emphasize single sinusoidal variation. Similarly, the decay parameter varies between a few tens to a few hundred days. As pointed out by \cite{angus18}, the flexibility of this model can easily lead to an overfitting of the data. To avoid this, we adopted a uniform prior distribution of the parameters (see, e.g., Table \ref{tab:earlygpfitparams}) that restricts the search range to realistic values. In addition, when the posterior distribution of the smoothing factor and decay time did not display a clear peak, we restricted the fit from six to five parameters by fixing the value of the smoothing factor at 0.7, which corresponds to the middle of the range of explored values and to the median value of stars with a constrained smoothing factor. We also explored a four-parameter fit in which both the smoothing factor and the decay time were fixed at 0.7 and 200 days, respectively. We found no significant differences with respect to the five-parameter fit. The value of the fixed decay time is a guess at best and may depend on the spectral type (early-M stars seem to have a shorter decay time, about 70 days): \citet{giles17} reported based on Kepler light curves that cooler stars have spots that last much longer, in particular for stars with longer rotational periods.

\begin{figure*}
     \centering
       \includegraphics[width=1.0\hsize]{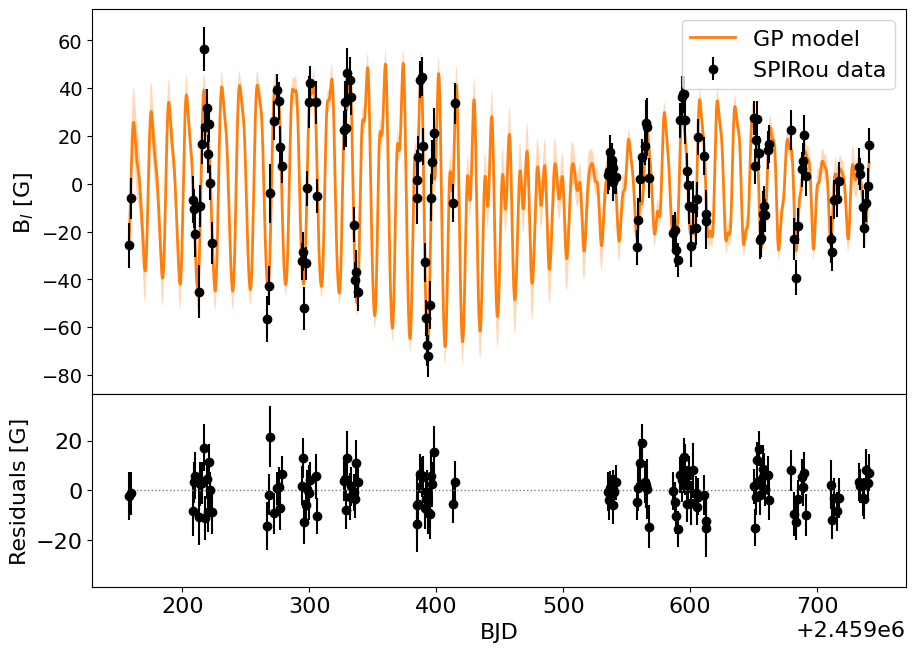}
      \caption{GP analysis of the SPIRou $B_\ell$ data of Gl\,410. In the top panel, the black points show the observed $B_\ell$ data, and the orange line shows the best-fit quasi-periodic GP model. The bottom panel shows the residuals with an RMS dispersion of 7.6~G.}
        \label{fig:GL410_blong_GP}
  \end{figure*}

\begin{figure*}
   \centering
       \includegraphics[width=1.0\hsize]{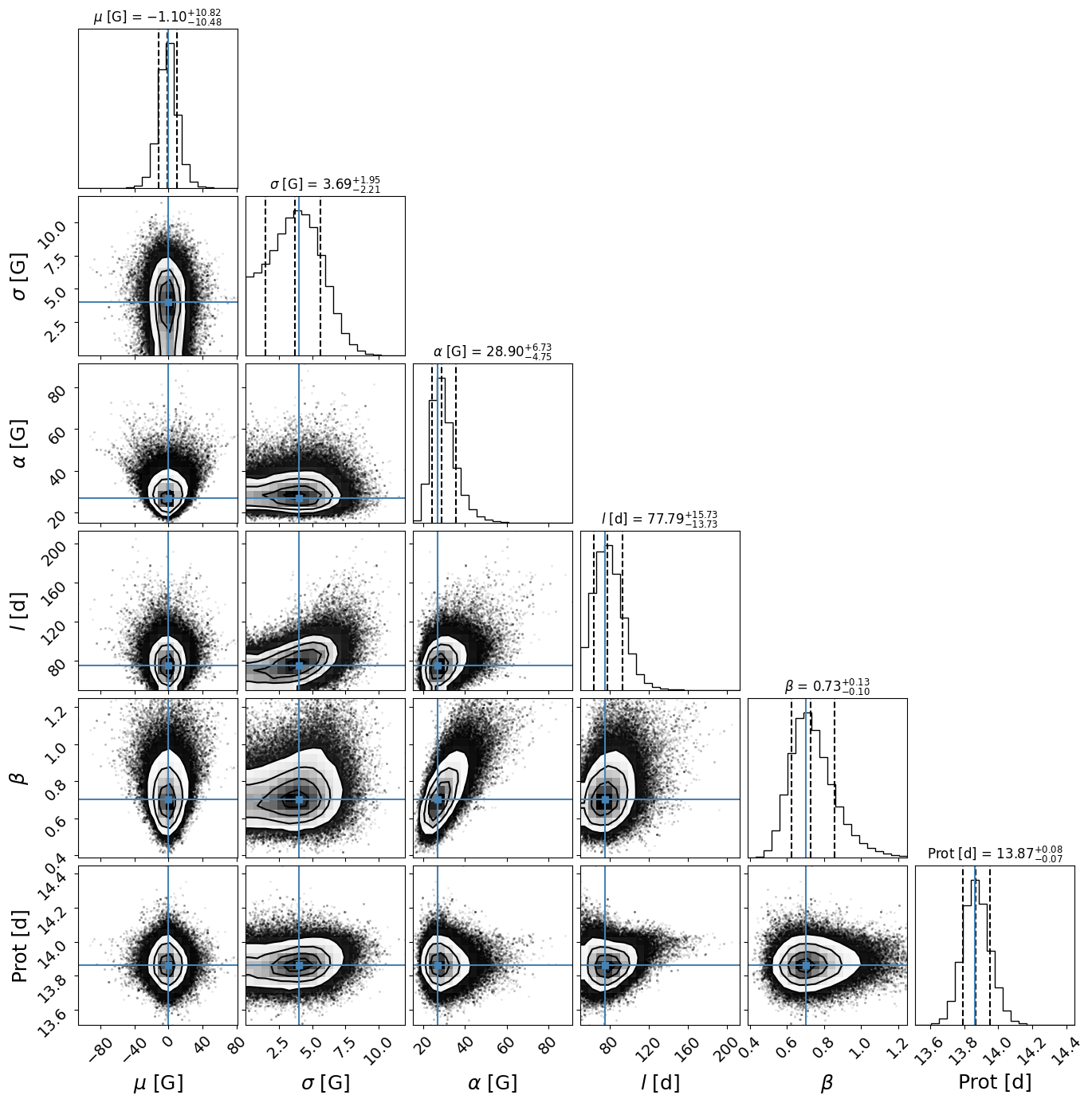}
      \caption{MCMC samples and the posterior distributions of parameters in the quasi-periodic GP analysis of the stellar activity in the SPIRou $B_\ell$ data of Gl\,410. The blue crosses mark the mode of the distribution, and the vertical dashed lines in the histograms indicate the median and the 16 and 84 percentiles of the posterior PDF. The shaded regions correspond to uncertainties of 1, 2, and 3 $\sigma$ in order of increasing radius.}
        \label{fig:GL410_blong_CP}
  \end{figure*}

\begin{figure*}
     \centering
       \includegraphics[width=1.0\hsize]{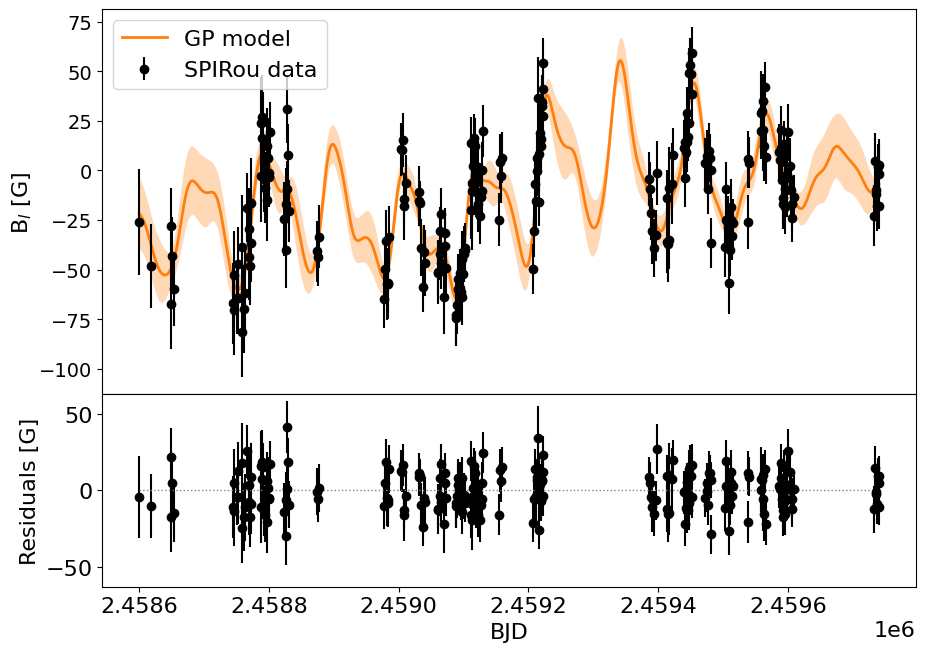}
      \caption{GP analysis of the SPIRou $B_\ell$ data of Gl\,905. In the top panel, the black points show the observed $B_\ell$ data, and the orange line shows the six-parameter best-fit quasi-periodic GP model. The bottom panel shows the residuals with an RMS dispersion of 13~G.}
        \label{fig:GL905_blong_GP}
  \end{figure*}

\begin{figure*}
   \centering
       \includegraphics[width=1.0\hsize]{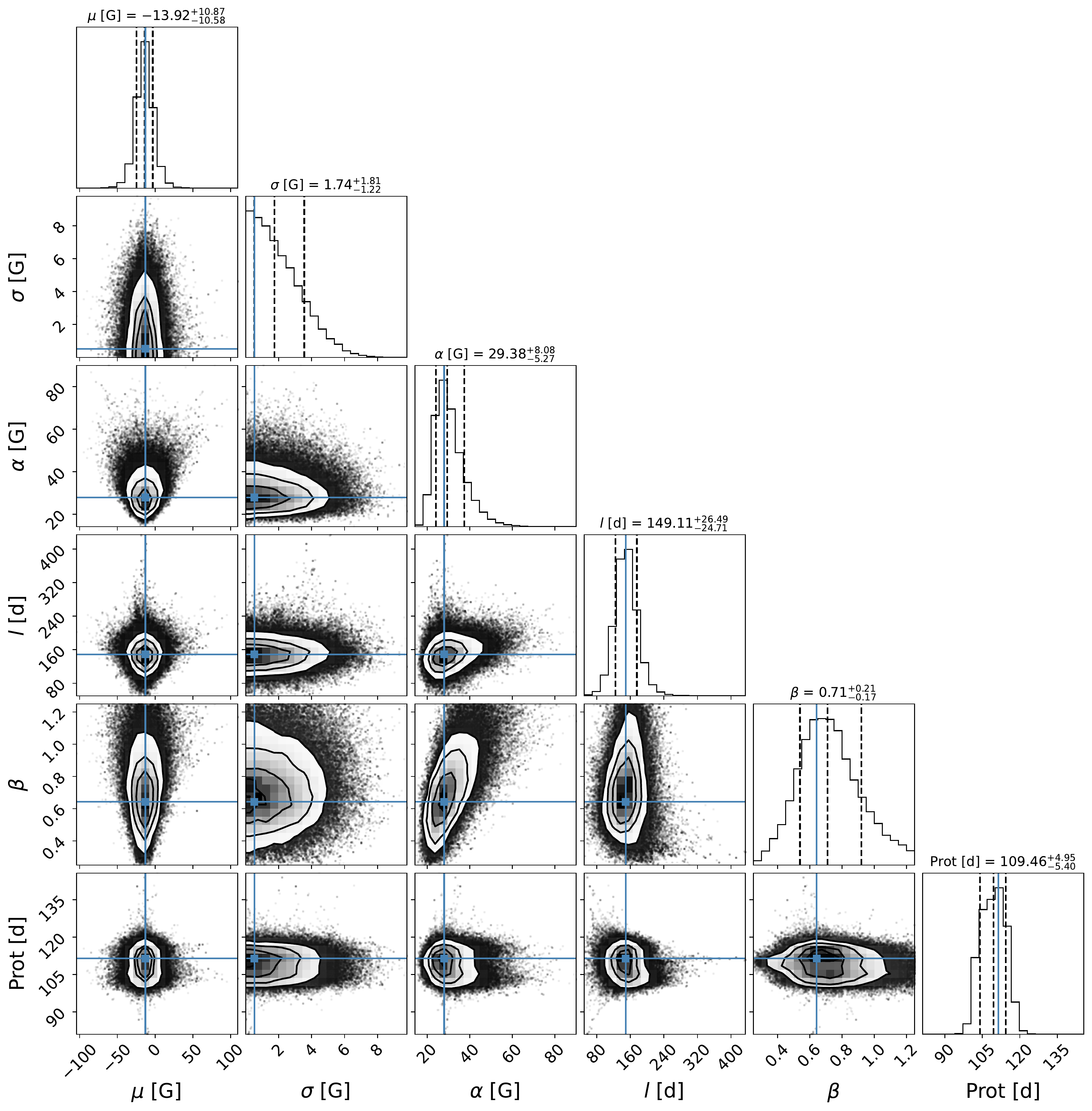}
      \caption{MCMC samples and the posterior distributions of the six-parameter fit of a quasi-periodic GP on the SPIRou $B_\ell$ data of Gl\,905. The same format is used as in Fig. \ref{fig:GL410_blong_CP}.}
        \label{fig:GL905_blong_CP}
  \end{figure*}

We used this GP framework to model the $B_\ell$ data, where we first fit the GP model parameters by maximizing the likelihood function defined in (Eq. 5.8)\footnote{http://gaussianprocess.org/gpml/chapters/RW5.pdf} of \citet{rasmussen06} and as implemented in \texttt{george}, and then we sampled the posterior distribution of the free parameters using a Bayesian Markov chain Monte Carlo (MCMC) framework with the package \texttt{emcee} \citep{foreman13}. We set the MCMC with 50 walkers, 1000 burn-in samples, and 5000 samples. The results of our analysis are illustrated in Fig. \ref{fig:GL410_blong_GP} and Fig. \ref{fig:GL905_blong_GP}, where we present the observed $B_\ell$ data and the best-fit GP model for Gl\,410 and Gl\,905, respectively. Fig. \ref{fig:GL410_blong_CP} and Fig. \ref{fig:GL905_blong_CP} show the MCMC samples and posterior distributions of the GP parameters for a six-parameter fit of the $B_\ell$ time series for Gl\,410 and Gl\,905, respectively.

\section{Results} \label{sec:results}

\subsection{Early-type M dwarfs}
\label{sec:early}

Our sample includes seven early-M dwarfs (M0V to M2V) with absolute magnitudes in the {\it Gaia} EDR3 $G$ band between 8 and 9. With masses between 0.5-0.6 \msun\ , they are partially convective. To compute the LSD profiles of these stars, we used a mask with atomic lines of the known Land\'e factor at 4000\,K and \logg=5.0. For six of them, the longitudinal magnetic field clearly varies, and accordingly, the rotation period is measured quite accurately. We used the six-parameter fit of the quasi-periodic GP model for them, but the decay time could not be well constrained for two of them. They are reported in Table \ref{tab:earlygpfitparams}, which lists the best-fit parameters measured as the median of the distribution, with uncertainties given by the 0.16 and 0.84 quantiles. Here and in Tables \ref{tab:middlegpfitparams} and \ref{tab:lategpfitparams}, the number of visits corresponds to the final number after poor polarimetric sequences were rejected, and it therefore differs from the numbers listed in Table \ref{tab:samplechars}. Results for Gl\,205 were already published in \citet{cortes23} based on a similar data set, but with the \texttt{Libre-Esprit} data reduction and analysis package: the agreement of the measured parameters is encouraging (see Table \ref{tab:gl205}). The rotation periods lie between 11 and 38~d for this sample. As the radius of early-type M dwarfs is in the range 0.5-0.6 \rsun, these rotation periods translate into equatorial velocities below 3 \kms. This \vsini \  is below the detectability threshold of our spectrograph (2-3 \kms). A broadening in the FWHM of the Stokes $I$ profile may be due to several factors when the rotation is negligible: for active stars, it may come from a Zeeman broadening effect, but for the quiet stars studied in this sample, it is probably due to other factors. In Sec. \ref{sec:conclusions} we return to this topic after the full sample is analyzed. For the earliest spectral-type star Gl\,338B, the Stokes V signal is clearly detected (see Fig. \ref{fig:gl338b}). However, we failed to detect any periodicity in the longitudinal magnetic field variation, probably due to the low number of visits (58 observations).

   \begin{table*}
    \caption[]{Best six-parameter fit of a quasi-periodic GP model obtained in our analysis of the stellar activity from the SPIRou $B_\ell$ data of Gl\,205, compared to the results from \citet{cortes23} using a similar data set, but the \texttt{Libre-Esprit} data reduction and analysis package. }
    \label{tab:gl205}
    \begin{tabular}{lccccccccc}
        \hline
        \noalign{\smallskip}
 Package & rotation period & mean $B_\ell$ & white noise & amplitude & decay time & smoothing factor & rms & $\chi^2_{\rm red}$ & visits \\
 & $P_{\rm rot}$ [d] & $\mu$ [G] & $\sigma$ [G] & $\alpha$ [G] & $l$ [d] & $\beta$ & [G] & & \\
        \noalign{\smallskip}
        \hline
        \noalign{\smallskip}
APERO & $34.3^{+0.4}_{-0.4}$ & $2.5^{+2.0}_{-2.0}$ & $1.0^{+0.5}_{-0.6}$ & $7.1^{+1.3}_{-1.0}$ & $67^{+15}_{-11}$ & $0.54^{+0.09}_{-0.07}$ & 2.5 & 0.79 & 152 \\
        \noalign{\smallskip}
Libre-Esprit & $34.4^{+0.5}_{-0.4}$ & $1.3^{+0.9}_{-0.9}$ & $0.4^{+0.2}_{-0.2}$ & $3.1^{+0.6}_{-0.4}$ & $63^{+13}_{-8}$ & $0.57^{+0.10}_{-0.08}$ & 0.94 & 0.84 & 153 \\
        \noalign{\smallskip}
        \hline
    \end{tabular}
  \end{table*}

The GP parameters appear to show that the decay time is pretty well constrained, with values clustering around 75~d. The smoothing parameter varies more widely, between 0.5 (several harmonics) and 1.0 (smooth variation). For two stars with poorly constrained smoothing parameter and decay time, we used a five-parameter fit by fixing the smoothing parameter to 0.7. The white noise is always compatible with 0 within $2 \sigma$, meaning that the GP does not need an additional variable to explain the dispersion, and the reduced \kid are close to 1. We reach an rms of the fit of a few G for these stars. The results are summarized in Table \ref{tab:earlygpfitparams}.

   \begin{table*}
    \caption[]{Best six-parameter fit of a quasi-periodic GP model obtained in our analysis of the stellar activity from the SPIRou $B_\ell$ data of the early-type M dwarfs in our sample. }
    \label{tab:earlygpfitparams}
    \begin{tabular}{lccccccccc}
        \hline
        \noalign{\smallskip}
 Star & rotation period & mean $B_\ell$ & white noise & amplitude & decay time & smoothing factor & rms & $\chi^2_{\rm red}$ & visits \\
 & $P_{\rm rot}$ [d] & $\mu$ [G] & $\sigma$ [G] & $\alpha$ [G] & $l$ [d] & $\beta$ & [G] & & \\
        \noalign{\smallskip}
 Priors & $\mathcal{U}(2,300)$ & $\mathcal{U}(-\infty,+\infty)$ & $\mathcal{U}(0,+\infty)$ & $\mathcal{U}(0,+\infty)$ & $\mathcal{U}(50,1000)$ & $\mathcal{U}(0.25,1.25)$ & & & \\
        \noalign{\smallskip}
        \hline
        \noalign{\smallskip}
Gl\,846 & $11.01^{+0.17}_{-0.22}$ & $-0.6^{+2.0}_{-2.1}$ & $2.6^{+1.0}_{-1.3}$ & $6.2^{+1.4}_{-1.2}$ & $57^{+18}_{-5}$ & $0.96^{+0.19}_{-0.22}$ & 6.1 & 1.05 & 188 \\
        \noalign{\smallskip}
Gl\,205 & $34.3^{+0.4}_{-0.4}$ & $2.5^{+2.0}_{-2.0}$ & $1.0^{+0.5}_{-0.6}$ & $7.1^{+1.3}_{-1.0}$ & $67^{+15}_{-11}$ & $0.54^{+0.09}_{-0.07}$ & 2.5 & 0.79 & 152 \\
        \noalign{\smallskip}
Gl\,410 & $13.87^{+0.08}_{-0.07}$ & $-1.1^{+10.8}_{-10.5}$ & $3.7^{+2.0}_{-2.2}$ & $28.9^{+6.7}_{-4.7}$ & $78^{+16}_{-14}$ & $0.73^{+0.13}_{-0.10}$ & 7.6 & 0.86 & 126 \\
        \noalign{\smallskip}
Gl\,880 & $37.7^{+0.8}_{-0.6}$ & $4.0^{+3.4}_{-3.1}$ & $0.8^{+0.8}_{-0.6}$ & $11.4^{+2.4}_{-1.7}$ & $94^{+24}_{-20}$ & $0.56^{+0.10}_{-0.08}$ & 3.8 & 0.65 & 162 \\
        \noalign{\smallskip}
Gl\,514 & $30.45^{+0.13}_{-0.14}$ & $-4.2^{+5.9}_{-5.4}$ & $2.1^{+1.3}_{-1.4}$ & $8.8^{+3.6}_{-2.4}$ & & $0.99^{+0.18}_{-0.25}$ & 8.2 & 1.10 & 165 \\
        \noalign{\smallskip}
Gl\,382 & $21.32^{+0.04}_{-0.03}$ & $-0.7^{+4.2}_{-4.2}$ & $2.8^{+1.2}_{-1.5}$ & $7.9^{+3.2}_{-1.9}$ & & $0.62^{+0.27}_{-0.18}$ & 6.7 & 1.11 & 114 \\
        \noalign{\smallskip}
        \hline
    \end{tabular}
  \end{table*}

Rotation periods were already known for all these stars, based on a variety of indicators: ZDI analysis, photometry (noted "phm" in Tables \ref{tab:earlylit}, \ref{tab:midlit}, and \ref{tab:latelit}), and activity indicators (noted "act" in the same tables). In Table \ref{tab:earlylit} we compare our new values to those from the literature. The agreement is good in general, but some literature values are discrepant for unclear reasons (see, e.g., Gl\,846) or reveal a harmonic (twice the frequency of rotation: Gl\,382 for \citet{sabotta21}). Photometry and activity measurements may be affected by sporadic variability (flares), and poor sampling may lead to differences or detection of harmonics of the true rotation period. A case-by-case discussion is difficult as the literature values are based on different techniques with varying precision, time sampling, and total time coverage. For our method, the time sampling and total coverage are generally adequate, and the precision is quite uniform after rejecting low S/N measurements.

   \begin{table*}
    \caption[]{Comparison of the rotation periods given in the literature with our measured values for the early-M stars.}
    \label{tab:earlylit}
    \begin{center}
    \scalebox{0.9}{
    \begin{tabular}{lccccccc}
        \hline
        \noalign{\smallskip}
 Reference & Category & Gl\,205 & Gl\,382 & Gl\,410 & Gl\,514 & Gl\,846 & Gl\,880 \\
        \noalign{\smallskip}
        \hline
        \noalign{\smallskip}
This work & $B_\ell$ & $34.3 \pm 0.4$ & $21.32 \pm 0.04$ & $13.87 \pm 0.08$ & $30.45 \pm 0.14$ & $11.01 \pm 0.20$ & $37.7 \pm 0.7$ \\
\citet{ks07} & phm & 33.61 & 21.56 & & & & \\
\citet{donati08} & ZDI & & & 14.0 & & & \\
\citet{bonfils13} & act & 32.8, 39.3 & & & & 10.7 & \\
\citet{suarez15} & act & $35.0 \pm 0.1$ & $21.7 \pm 0.1$ & & $28.0 \pm 2.9$ & $31.0 \pm 0.1$ & $37.5 \pm 0.1$ \\
\citet{suarez16} & phm & $33.4 \pm 0.1$ & $21.2 \pm 0.1$ & & & & \\
\citet{hebrard16} & ZDI & $33.63 \pm 0.37$ & & $13.83 \pm 0.10$ & & $10.73 \pm 0.10$ & \\
\citet{suarez17} & act & $34.8 \pm 1.3$ & $21.8 \pm 0.1$ & & $30.0 \pm 0.9$ & $26.3 \pm 5.6$ & $37.2 \pm 6.7$ \\
\citet{diez19} & phm & $33.8 \pm 0.6$ & $21.6 \pm 0.2$ & $14.6 \pm 0.2$ & & $29.5 \pm 0.1$ & $39.5 \pm 0.2$ \\
\citet{sabotta21} & act & 37.08 & 10.65, 21.4 & & & & \\

        \hline
    \end{tabular}}
    \end{center}
  \end{table*}

\subsection{Mid-type M dwarfs}
\label{sec:middle}

The most numerous subcategory in our sample comprises 26 stars with \MG~ between 9 and 11, spectral types ranging from M1V to M4V, and masses between 0.2 and 0.5 \msun. This group crosses the so-called Jao gap \citep{jao18} at about \MG=10.2, \BP-\RP=2.3, and includes the transition between partially convective and fully convective M dwarfs. As the transition mass between these two regimes is poorly defined (0.2 to 0.35\,\msun) and depends on metallicity (see \citealt{feiden21} for a more detailed discussion), we preferred not to try to define a finer grid of magnitudes or spectral types with the risk of a possibly inhomogeneous subcategory. To compute the LSD profiles of these stars, we used a mask with atomic lines of the known Land\'e factor at 3500\,K and \logg=5.0.

In contrast with the group of early-M dwarfs, here only half of the stars have a detected rotation period. We first tried a six-parameter fit of the quasi-periodic GP model for them, but only three stars have a well-constrained decay time at about 100\,days. For the other stars for which we were able to determine a rotation period, we use a five or even a four parameters fit, fixing the smoothing parameter at 0.7 (five-parameter fit), and when necessary, the decay time to 200\,days (four-parameter fit). We detect a very long periodic variation of about 450\,days for Gl\,411 in this group, which clearly disagrees with the shorter period reported from photometry by \citet{diaz19} of $56.15 \pm 0.27$d, and is unexpectedly long compared to all the M dwarfs with known stellar rotation periods (see Fig.~\ref{fig:protmassdiagram}). It would imply an unknown mechanism of angular momentum loss. We show the GP fit in Fig.~\ref{fig:GL411_blong_GP} that corresponds to this particular star, and its associated corner plot is presented in Fig.~\ref{fig:GL411_blong_CP}. The measured period may correspond to a cyclic variation in the magnetic field that is more related to the variation of the activity than to the stellar rotation.

We were able to measure the rotation period for half of the mid-M dwarfs compared to 100\% for the early-M dwarfs. Furthermore, a six-parameter fit could only be measured for three stars compared to four out of six for the early-M dwarfs. This may come from a longer decay time that is not constrained enough by our three-year survey. The shape of the variation in the longitudinal magnetic field varies strongly, from almost sinusoidal variations for Gl\,169.1A (smoothing factor of 1) to a variation featuring only a few harmonics for Gl\,48, GJ\,3378 or GJ\,4333 (smoothing factor of 0.6). The amplitude of the variations ranges from very weak (5~G for Gl\,411) to about 20 G for GJ\,4333 and Gl\,876. For this group, the white-noise component of the GP is always compatible with 0 at $2 \sigma$, as was the case for the early-M dwarfs.

   \begin{table*}
    \caption[]{Best six to four parameters fit of a quasi-periodic GP model obtained in our analysis of the stellar activity from the SPIRou $B_\ell$ data of the mid-range M dwarfs in our sample. The asterisk after a star name means that the measured period is more uncertain and was obtained by a five or four parameters fit only.}
    \begin{center}
    \label{tab:middlegpfitparams}
    \scalebox{0.9}{
    \begin{tabular}{lccccccccc}
        \hline
        \noalign{\smallskip}
 Star & rotation period & mean $B_\ell$ & white noise & amplitude & decay time & smoothing factor & rms & $\chi^2_{\rm red}$ & visits \\
 & $P_{\rm rot}$ [d] & $\mu$ [G] & $\sigma$ [G] & $\alpha$ [G] & $l$ [d] & $\beta$ & [G] & & \\
        \noalign{\smallskip}
 Priors & $\mathcal{U}(2,300)$ & $\mathcal{U}(-\infty,+\infty)$ & $\mathcal{U}(0,+\infty)$ & $\mathcal{U}(0,+\infty)$ & $\mathcal{U}(50,1000)$ & $\mathcal{U}(0.25,1.25)$ & & & \\
        \noalign{\smallskip}
        \hline
        \noalign{\smallskip}
Gl\,752A & $53.2^{+5.5}_{-3.0}$ & $0.2^{+3.4}_{-3.2}$ & $1.6^{+1.3}_{-1.1}$ & $8.6^{+2.9}_{-1.9}$ & $93^{+55}_{-29}$ & $0.89^{+0.24}_{-0.29}$ & 6.2 & 0.87 & 128 \\
        \noalign{\smallskip}
Gl\,48 & $51.2^{+1.4}_{-1.4}$ & $-7.0^{+3.6}_{-3.7}$ & $1.8^{+1.6}_{-1.2}$ & $10.2^{+3.2}_{-2.1}$ & $112^{+55}_{-36}$ & $0.62^{+0.22}_{-0.16}$ & 9.4 & 0.86 & 188 \\
        \noalign{\smallskip}
Gl\,15A & $44.3^{+2.0}_{-2.0}$ & $-1.0^{+1.8}_{-1.7}$ & $1.0^{+0.9}_{-0.7}$ & $5.4^{+1.4}_{-1.0}$ & $85^{+36}_{-22}$ & $0.71^{+0.28}_{-0.23}$ & 5.6 & 0.82 & 246 \\
        \noalign{\smallskip}
Gl\,849 & $41.4^{+0.4}_{-0.4}$ & $4.7^{+6.3}_{-5.9}$ & $2.4^{+1.6}_{-1.6}$ & $12.1^{+5.1}_{-3.2}$ & & $0.87^{+0.24}_{-0.23}$ & 10.0 & 1.01 & 185 \\
        \noalign{\smallskip}
Gl\,411 & $471^{+41}_{-40}$ & $5.9^{+2.4}_{-2.1}$ & $0.8^{+0.8}_{-0.6}$ & $4.2^{+1.8}_{-1.1}$ & & $0.68^{+0.32}_{-0.27}$ & 6.3 & 0.94 & 212 \\
        \noalign{\smallskip}
Gl\,687 & $56.5^{+1.3}_{-0.5}$ & $4.9^{+8.9}_{-6.4}$ & $1.7^{+1.3}_{-1.1}$ & $13.3^{+7.3}_{-4.3}$ & & $0.97^{+0.19}_{-0.26}$ & 8.2 & 1.10 & 214 \\
        \noalign{\smallskip}
Gl\,725A & $103.5^{+4.6}_{-5.1}$ & $-14.9^{+3.4}_{-3.1}$ & $1.2^{+1.2}_{-0.8}$ & $8.7^{+2.4}_{-1.6}$ & & & 8.1 & 0.81 & 214 \\
        \noalign{\smallskip}
Gl\,251$^{\ast}$ & $98.7^{+11.5}_{-4.8}$ & $17.4^{+3.6}_{-3.6}$ & $1.7^{+1.6}_{-1.2}$ & $8.3^{+2.4}_{-1.9}$ & & & 9.9 & 0.85 & 177 \\
        \noalign{\smallskip}
GJ\,4333$^{\ast}$ & $72.0^{+0.9}_{-1.2}$ & $6.8^{+8.9}_{-8.0}$ & $2.6^{+2.1}_{-1.7}$ & $18.5^{+8.9}_{-4.1}$ & & $0.55^{+0.25}_{-0.17}$ & 12.7 & 0.83 & 190 \\
        \noalign{\smallskip}
Gl\,876 & $82.8^{+2.0}_{-0.7}$ & $3.5^{+13.3}_{-13.5}$ & $2.0^{+1.9}_{-1.4}$ & $23.0^{+12.3}_{-7.1}$ & & $0.72^{+0.24}_{-0.19}$ & 8.5 & 0.92 & 88 \\
        \noalign{\smallskip}
PM J09553-2715$^{\ast}$ & $70.5^{+5.7}_{-1.9}$ & $15.4^{+9.8}_{-11.7}$ & $2.7^{+2.4}_{-1.9}$ & $15.1^{+8.7}_{-5.1}$ & & $0.82^{+0.30}_{-0.33}$ & 11.8 & 0.91 & 74 \\
        \noalign{\smallskip}
GJ\,3378$^{\ast}$ & $92.1^{+4.1}_{-5.3}$ & $13.6^{+5.6}_{-5.4}$ & $4.6^{+2.3}_{-2.7}$ & $12.0^{+5.1}_{-2.8}$ & & $0.61^{+0.32}_{-0.23}$ & 13.9 & 1.00 & 174 \\
        \noalign{\smallskip}
Gl\,169.1A & $91.9^{+4.1}_{-2.6}$ & $2.3^{+8.4}_{-7.2}$ & $1.9^{+1.9}_{-1.3}$ & $11.4^{+5.3}_{-3.3}$ & & $1.04^{+0.15}_{-0.23}$ & 13.4 & 0.91 & 172 \\
        \noalign{\smallskip}
        \hline
    \end{tabular}}
    \end{center}
  \end{table*}

In Table \ref{tab:midlit} we compare our new values of the rotation period to those from the literature. The agreement is good in general, but some literature values are discrepant. Excluding possible sampling problems or a low number of measurements in one case (Gl\,876), there may be more fundamental reasons for some techniques to succeed or fail on a given star: photometric variations may be affected by flares, while the longitudinal magnetic field of the star may stay constant due to an axisymmetric field topology that prevents us from detecting a periodic variation even when the Stokes $V$ profiles show a clear detection. At least in one case, Gl\,411, the very long measured period may in fact reflect a cyclic variation of activity rather than the stellar rotation period. We also show at the end of Sec. \ref{sec:late} that some stars have a clear detection of the Stokes $V$ profile, but no rotation period detected when the magnetic field is axisymmetric.

   \begin{table*}
    \caption[]{Comparison of the rotation periods given in the literature with our measured values for the mid-M stars.}
    \label{tab:midlit}
    \begin{center}
    \scalebox{0.7}{
    \begin{tabular}{lcccccccccccc}
        \hline
        \noalign{\smallskip}
 Reference & Category & Gl\,752A & Gl\,48 & Gl\,15A & Gl\,849 & Gl\,411 & Gl\,687 & Gl\,251 & GJ\,4333 & Gl\,876 & GJ\,3378 \\
        \noalign{\smallskip}
        \hline
        \noalign{\smallskip}
This work & $B_\ell$ & $53 \pm 4$ & $51.2 \pm 1.4$ & $44.3 \pm 2.0$ & $41.4 \pm 0.4$ & $470 \pm 40$ & $56.5 \pm 0.9$ & $99 \pm 8$ & $72.0 \pm 1.0$ & $82.8 \pm 1.4$ & $92 \pm 5$ \\
\citet{rivera05} & phm & & & & & & & & & $96.7 \pm 1.0$ \\
\citet{bonfils13} & act & & & & 2000 ? & & & & & 61.0, 30.1 \\
\citet{burt14} & phm & & & & & & $61.8 \pm 1.0$ \\
\citet{howard14} & phm & & & $43.82 \pm 0.56$ & & & & & & & \\
\citet{suarez15} & act & $46.5 \pm 0.3$ & & & $39.2 \pm 6.3$ & & & & & $87.3 \pm 5.7$ \\
\citet{nelson16} & act & & & & & & & & & $95 \pm 1$ \\
\citet{suarez16} & phm & $46.0 \pm 0.2$ & & & & & & \\
\citet{moutou17} & ZDI & & & & & & & $90 \pm 10$ & & & \\
\citet{suarez17} & act & & & & & & & & & $90.9 \pm 16.5$ \\
\citet{suarez18} & act & & & $45.0 \pm 4.4$ & & & & \\
\citet{diaz19} & phm & & & & & $56.15 \pm 0.27$ & & \\
\citet{diez19} & phm & $46.0 \pm 0.2$ & $51.5 \pm 2.6$ & & & & & $18.1 \pm 0.3$ & $74.7 \pm 0.7$ & $81.0 \pm 0.8$ & \\
\citet{reinhold20} & phm & & & & & & & & & $31.31 \pm 8.15$ & \\
\citet{stock20} & phm & & & & & & & $122.1 \pm 2.2$ & & & \\
\citet{sabotta21} & act & 174.48 & 43.39 & & & & & 67.59, 119.48 & & & 83.39 \\
        \hline
    \end{tabular}}
    \end{center}
  \end{table*}

We also tested the results of the \texttt{Libre-Esprit} pipeline as given in Carmona et al. (in prep) for Gl\,388 (AD\,Leo), even though this star was discarded from our sample because it is too active. Table \ref{tab:gl388} compares the results, which agree excellently well.

   \begin{table*}
    \caption[]{Best six-parameter fit of a quasi-periodic GP model obtained in our analysis of the stellar activity from the SPIRou $B_\ell$ data of Gl\,388 (AD\,Leo), compared to the results from Carmona et al. (in prep) using a similar data set, but the \texttt{Libre-Esprit} data reduction and analysis package. }
    \label{tab:gl388}
    \begin{tabular}{lccccccccc}
        \hline
        \noalign{\smallskip}
 Package & rotation period & mean $B_\ell$ & white noise & amplitude & decay time & smoothing factor & rms & $\chi^2_{\rm red}$ & visits \\
 & $P_{\rm rot}$ [d] & $\mu$ [G] & $\sigma$ [G] & $\alpha$ [G] & $l$ [d] & $\beta$ & [G] & & \\
        \noalign{\smallskip}
        \hline
        \noalign{\smallskip}
APERO & $2.2301^{+0.0019}_{-0.0017}$ & $-156^{+43}_{-42}$ & $4.6^{+3.3}_{-3.0}$ & $68^{+21}_{-14}$ & $220^{+41}_{-40}$ & $1.35^{+0.11}_{-0.19}$ & 12.7 & 0.91 & 70 \\
        \noalign{\smallskip}
Libre-Esprit & $2.2304^{+0.0015}_{-0.0013}$ & $-156^{+47}_{-44}$ & $5.4^{+3.7}_{-3.6}$ & $67^{+23}_{-15}$ & $255^{+96}_{-77}$ & $1.36^{+0.10}_{-0.17}$ & 14.7 & 0.96 & 69 \\
        \noalign{\smallskip}
        \hline
    \end{tabular}
  \end{table*}

\subsection{Late-M dwarfs}
\label{sec:late}
The last group contains ten stars with $M_G$ above 11 (up to 13.5), spectral types from M3.5V to M5.5V, and masses between 0.1 and 0.2 \msun. These stars are fully convective. Stars with later spectral types (M6.0V and M6.5V) in the WP1 sample are all active stars with short rotation periods and are therefore excluded from this study. To compute the LSD profiles of these late-type stars, we used a mask with atomic lines of the known Land\'e factor at 3000~K and \logg=5.0. We generally had to use a four-parameter fit of the quasi-periodic GP model for them, fixing the smoothing parameter to 0.7 and the decay time to 200 days, except for two stars in this group, for which a six-parameter fit gives well-constrained values of the smoothing factor and the decay time. The others are marked with an asterisk in Table \ref{tab:lategpfitparams}. For only one star, Gl\,447 (Ross\,128, FI\,Vir), are we unable to confirm the long rotation period measured from ASAS photometry by \cite{suarez16} ($165.1 \pm 0.8$~d) and \citet{diez19} ($163 \pm 3$~d). This is probably due to the low number and distribution of the visits (57, clustered into two groups) compared to the other late-M dwarfs, which prevents us from determining the rotation period of Gl\,447. This highlights once again that our set limit to 50 visits with adequate sampling is essential. Below this value, the determination of the rotation period becomes difficult.

   \begin{table*}
    \caption[]{Best six, five, or four parameters fit of a quasi-periodic GP model obtained in our analysis of the stellar activity from the SPIRou $B_\ell$ data of the late-M dwarfs in our sample. The asterisk after a star name means that the measured period is more uncertain and was obtained by a four-parameter fit only.}
    \label{tab:lategpfitparams}
    \begin{center}
    \begin{tabular}{lccccccccc}
        \hline
        \noalign{\smallskip}
 Star & rotation period & mean $B_\ell$ & white noise & amplitude & decay time & smoothing factor & rms & $\chi^2_{\rm red}$ & visits \\
 & $P_{\rm rot}$ [d] & $\mu$ [G] & $\sigma$ [G] & $\alpha$ [G] & $l$ [d] & $\beta$ & [G] & & \\
        \noalign{\smallskip}
 Priors & $\mathcal{U}(2,300)$ & $\mathcal{U}(-\infty,+\infty)$ & $\mathcal{U}(0,+\infty)$ & $\mathcal{U}(0,+\infty)$ & $\mathcal{U}(50,1000)$ & $\mathcal{U}(0.25,1.25)$ & & & \\
        \noalign{\smallskip}
        \hline
        \noalign{\smallskip}
GJ\,1289 & $74.0^{+1.5}_{-1.3}$ & $47^{+25}_{-25}$ & $2.4^{+2.5}_{-1.7}$ & $67^{+21}_{-12}$ & $142^{+33}_{-26}$ & $0.64^{+0.17}_{-0.11}$ & 14.7 & 0.67 & 180 \\
        \noalign{\smallskip}
GJ\,1103$^{\ast}$ & $139^{+22}_{-23}$ & $8.2^{+10.1}_{-10.1}$ & $4.1^{+4.0}_{-2.9}$ & $18.2^{+8.9}_{-5.9}$ & & & 16.4 & 0.83 & 62 \\
        \noalign{\smallskip}
Gl\,699$^{\ast}$ & $137.1^{+6.8}_{-4.0}$ & $3.0^{+5.7}_{-5.8}$ & $1.1^{+1.1}_{-0.8}$ & $14.7^{+3.0}_{-2.4}$ & & & 8.7 & 0.77 & 243 \\
        \noalign{\smallskip}
Gl\,15B$^{\ast}$ & $116.5^{+6.4}_{-4.7}$ & $-0.3^{+6.3}_{-6.2}$ & $2.1^{+2.1}_{-1.4}$ & $13.8^{+3.8}_{-2.8}$ & & & 16.5 & 0.78 & 184 \\
        \noalign{\smallskip}
GJ\,1151$^{\ast}$ & $158^{+14}_{-9}$ & $-9.6^{+13.4}_{-13.2}$ & $2.7^{+2.6}_{-1.9}$ & $29.6^{+8.2}_{-5.5}$ & & & 15.7 & 0.83 & 153 \\
        \noalign{\smallskip}
Gl\,905 & $109.5^{+4.9}_{-5.4}$ & $-14^{+11}_{-11}$ & $1.7^{+1.8}_{-1.2}$ & $29.4^{+8.1}_{-5.3}$ & $149^{+26}_{-25}$ & $0.71^{+0.21}_{-0.17}$ & 12.6 & 0.73 & 216 \\
        \noalign{\smallskip}
GJ\,1286$^{\ast}$ & $203^{+14}_{-21}$ & $36^{+19}_{-20}$ & $5.7^{+4.5}_{-3.8}$ & $44^{+14}_{-10}$ & & & 25.0 & 0.89 & 108 \\
        \noalign{\smallskip}
GJ\,1002 & $93.0^{+1.4}_{-1.7}$ & $-6.4^{+12.5}_{-14.2}$ & $4.1^{+3.4}_{-2.8}$ & $20.4^{+9.7}_{-5.9}$ & & $0.90^{+0.24}_{-0.31}$ & 19.9 & 1.05 & 154 \\
        \noalign{\smallskip}
        \hline
    \end{tabular}
    \end{center}
  \end{table*}

In Table \ref{tab:latelit} we compare our new rotation periods to those from the literature. The agreement is not as good as for earlier M dwarfs, probably because their rotation periods are longer and the uncertainties are larger. It becomes naturally harder to determine very long periods as we are limited by the time range of the SPIRou observations.

   \begin{table*}
    \caption[]{Comparison of the rotation periods given in the literature with our measured values for the late Ms.}
    \label{tab:latelit}
    \begin{center}
    \begin{tabular}{lccccccc}
        \hline
        \noalign{\smallskip}
 Reference & Category & GJ\,1289 & Gl\,699 & GJ\,1151 & Gl\,905 & GJ\,1286 & GJ\,1002 \\
        \noalign{\smallskip}
        \hline
        \noalign{\smallskip}
This work & $B_\ell$ & $74.0 \pm 1.4$ & $137 \pm 5$ & $158 \pm 12$ & $110 \pm 5$ & $203 \pm 18$ & $93.0 \pm 1.6$ \\
\citet{benedict98} & phm & & 130.4 & & & & \\
\citet{irwin11} & phm & & & 132 & & & \\
\citet{suarez15} & act & & $148.6 \pm 0.1$ & & & & \\
\citet{moutou17} & ZDI & $54 \pm 4$ & & & & & \\
\citet{newton18} & phm & & & & & 88.92 & \\
\citet{toledo19} & phm & & $145 \pm 15$ & & & & \\
\citet{diez19} & phm & $83.6 \pm 7.0$ & & $125 \pm 23$ & $106 \pm 6$ & & \\
\citet{sabotta21} & act & & 311.25 & & 178.74 & & \\
\citet{suarez22} & act & & & & & & $126 \pm 15$ \\
        \hline
    \end{tabular}
    \end{center}
  \end{table*}

Finally, in Table \ref{tab:undetectedgpfitparams}, we present the data of the 16 stars for which we were unable to measure a rotation period. For these stars, we arbitrarily fixed the decay parameter to 200~d and the smoothing factor to 0.7 to obtain some indications about the mean $B_\ell$, residual white noise, and amplitude of the field. The GP also gives a value of the rotation period, but we do not list it as it is not well constrained. We assumed that the values of the other parameters of the GP still bear some information, although the fact that the rotation period is not constrained and that two parameters of the GP are arbitrarily fixed limits the value of this information. For completeness, we list a period and its reference, when available in the literature, but this information was not used in our GP fit. 

The Stokes $V$ profiles in Appendix \ref{sec:stokesV} may tentatively explain these nondetections by their magnetic topology: if the magnetic field is axisymmetric, we cannot detect the rotation modulation. To test this interpretation, we measured the rate of detection of Stokes V profiles for a given star by comparison to the noise measured in a velocity region well outside of the line. We find that this metric quantifies the visual impression of the time series of Stokes V profiles quite well, as shown in Appendix \ref{sec:stokesV}. Then, we subtracted a median Stokes V profile for each star and measured the detection rate again. For some stars, it changes dramatically, and we tentatively interpret this change by the subtraction of a constant component due to an axisymmetric magnetic field. This is the case of Gl\,408, Gl\,338B, Gl\,436, GJ\,4063, and Gl\,617B. For other stars, the detection rate does not significantly change when the median-subtracted profiles are compared to the original Stokes V profiles. This is the case of PM\,J21463+3813, Gl\,412A, PM\,J08402+3127, Gl\,480, and GJ\,1148 among the stars without a detected rotation period. We plan a more detailed study of the magnetic topology of the stars in our sample to further investigate the reasons for these nondetections. 

   \begin{table*}
    \caption[]{Best four-parameter fit of a quasi-periodic GP model obtained in our analysis of the stellar activity from the SPIRou $B_\ell$ data of the M dwarfs in our sample without a clear periodic variation detection.}
    \label{tab:undetectedgpfitparams}
    \begin{center}
    \scalebox{0.8}{
    \begin{tabular}{lcccccccc}
        \hline
        \noalign{\smallskip}
 Star & mean $B_\ell$ & white noise & amplitude & rms & $\chi^2_{\rm red}$ & number of visits & rotation period from literature & reference \\
 & $\mu$ [G] & $\sigma$ [G] & $\alpha$ [G] & [G] & & & $P_{\rm rot}$ [d]  & \\
        \noalign{\smallskip}
 Priors & $\mathcal{U}(-\infty,+\infty)$ & $\mathcal{U}(0,+\infty)$ & $\mathcal{U}(0,+\infty)$ & & & & & \\
        \noalign{\smallskip}
        \hline
        \noalign{\smallskip}
Gl\,338B & $-8.1^{+1.6}_{-1.6}$ & $1.3^{+1.0}_{-0.9}$ & $3.1^{+1.6}_{-1.4}$ & 4.4 & 1.03 & 58 & 16.66 & \citet{sabotta21} \\
        \noalign{\smallskip}
Gl\,617B & $19.3^{+4.0}_{-4.0}$ & $1.4^{+1.5}_{-1.0}$ & $8.9^{+2.9}_{-2.1}$ & 8.2 & 0.71 & 142 & \\
        \noalign{\smallskip}
Gl\,412A & $13.7^{+4.1}_{-4.3}$ & $4.0^{+1.2}_{-1.4}$ & $10.7^{+3.5}_{-2.5}$ & 9.7 & 1.23 & 174 & $100.9 \pm 0.3$ & \citet{suarez18} \\
        \noalign{\smallskip}
Gl\,480 & $6.6^{+3.5}_{-3.3}$ & $5.5^{+1.8}_{-2.1}$ & $6.9^{+2.8}_{-2.3}$ & 11.0 & 1.28 & 104 & & \\
        \noalign{\smallskip}
Gl\,436 & $-10.4^{+2.2}_{-2.0}$ & $1.6^{+1.6}_{-1.1}$ & $4.4^{+2.1}_{-1.7}$ & 7.5 & 0.81 & 85 & $44.09 \pm 0.08$ & \citet{bourrier18} \\
        \noalign{\smallskip}
Gl\,408 & $-43.8^{+3.6}_{-3.6}$ & $1.4^{+1.4}_{-1.0}$ & $8.7^{+2.6}_{-2.0}$ & 9.1 & 0.79 & 168 & & \\
        \noalign{\smallskip}
Gl\,317 & $-10.8^{+8.4}_{-8.8}$ & $2.7^{+2.5}_{-1.9}$ & $14.5^{+8.2}_{-4.1}$ & 11.9 & 0.97 & 76 & & \\
        \noalign{\smallskip}
GJ\,4063 & $17.8^{+3.3}_{-3.2}$ & $2.9^{+1.9}_{-1.9}$ & $7.1^{+2.3}_{-1.8}$ & 11.4 & 0.96 & 204 & $40.2 \pm 0.8$ & \citet{diez19} \\
        \noalign{\smallskip}
GJ\,1012 & $6.4^{+2.6}_{-2.6}$ & $3.2^{+2.7}_{-2.2}$ & $4.0^{+3.8}_{-2.7}$ & 15.9 & 0.99 & 135 & & \\
        \noalign{\smallskip}
GJ\,1148 & $-9.1^{+3.9}_{-3.5}$ & $2.3^{+2.4}_{-1.6}$ & $7.1^{+4.1}_{-2.9}$ & 13.9 & 0.81 & 101 & $71.5 \pm 5.1$ & \citet{diez19} \\
        \noalign{\smallskip}
PM J08402+3127 & $16^{+20}_{-21}$ & $4.6^{+3.8}_{-3.1}$ & $39^{+16}_{-9}$ & 18.8 & 0.98 & 139 & $118 \pm 14$ & \citet{diez19} \\
        \noalign{\smallskip}
Gl\,725B & $-4.6^{+5.2}_{-5.1}$ & $1.2^{+1.3}_{-0.8}$ & $12.1^{+3.7}_{-2.8}$ & 10.7 & 0.75 & 209 \\
        \noalign{\smallskip}
GJ\,1105 & $1.2^{+3.6}_{-3.1}$ & $2.4^{+2.2}_{-1.7}$ & $7.3^{+3.2}_{-2.7}$ & 13.5 & 0.93 & 161 \\
        \noalign{\smallskip}
Gl\,445 & $-1.8^{+2.7}_{-2.6}$ & $3.4^{+3.2}_{-2.4}$ & $3.2^{+3.7}_{-2.3}$ & 17.4 & 0.97 & 90 & & \\
        \noalign{\smallskip}
PM J21463+3813 & $-0.4^{+7.5}_{-7.7}$ & $3.1^{+3.1}_{-2.2}$ & $16.3^{+6.6}_{-4.6}$ & 20.7 & 0.84 & 176 \\
        \noalign{\smallskip}
Gl\,447 & $20.5^{+9.1}_{-9.5}$ & $4.8^{+3.0}_{-3.0}$ & $16.6^{+8.5}_{-5.3}$ & 14.9 & 1.13 & 57 & $165.1 \pm 0.8$ & \citet{suarez16} \\
        \noalign{\smallskip}
        \hline
    \end{tabular}}
    \end{center}
  \end{table*}

\section{Discussion and conclusions}
\label{sec:conclusions}

We have shown that spectropolarimetry is a useful technique for measuring the rotation period of a star, even for quiet M dwarfs. In our sample of 43 such stars, we were able to reliably measure the rotation period for 27 stars, 8 of which were previously unknown. The rotation periods cover a wide range in this sample of quiet stars, from 10 to 450 days. The agreement with other techniques, such as photometry or stellar activity indicators is good, except for a few stars (e.g., Gl\,251, Gl\,411, and Gl\,846) for which some techniques find a harmonic of the rotation period or converge toward very different values. The amplitude of variation in the longitudinal magnetic field for these quiet M dwarfs ranges from 3 G (Gl\,338B, Gl\,445) to 20 G for the majority of the stars in our sample, but it can reach up to 70 G in some cases (GJ\,1289), which is comparable to more active stars such as AD\,Leo (Gl\,388: $45 \pm 6$\;G, Carmona et al., in prep) or EV\,Lac (Gl\,873: $149 \pm 17$\;G) over the same period of observations. The FWHM of the median Stokes $I$ profile is another important measurement, as it may reveal a broadening due to the Zeeman effect for active stars. As the FWHM is sensitive to the total magnetic field (small and large scale),while the longitudinal magnetic field is more sensitive to the large-scale field, a large amplitude of variation in the longitudinal field does not necessarily correlate with a high FWHM, and both measures are therefore important. The average value over our sample is 6.07 \kms , with a standard deviation of 0.66 \kms. This is clearly larger than the mean uncertainty of each FWHM in Table \ref{tab:samplechars}. This shows a real dispersion among the mean FWHM values, which will be further investigated by using different line lists (low and high Land\'e factors) and by studying the time variation in the measured FWHM for a given star, as in Bellotti et al. (in prep) for AD\,Leo.

We divided our sample into three sub-categories according to the absolute magnitude in {\it Gaia} $G$ band, roughly corresponding to early-, mid- and late-M dwarfs. There is a clear tendency to lower values of the FWHM of the median Stokes $I$ profile toward mid-type stars, with averages and standard deviations for the three groups of 6.65 and 0.59 \kms (early type), 5.73 and 0.34 \kms (mid type), and 6.49 and 0.75 \kms (late type), respectively. The larger dispersion in the early-type and late-type groups may be due to a few active outliers: removing Gl\,410 from the early-type group gives new average and standard deviation of 6.45 and 0.40 \kms, respectively, while for the late-type group, removing GJ\,1286 and GJ\,1289 gives 6.14 and 0.52 \kms, respectively, now more compatible with the mid-type group, given the small size of the early- and late-type groups. The most extreme values correspond to Gl\,411 ($4.93\pm0.18$ \kms) and Gl\,410 ($7.73 \pm 0.21$ \kms). A test on GJ\,1289 shows that the measured FWHM clearly depends on the minimum adopted depth (0.03 in this work) and the LSD mask (3000 and 3500\,K give different results for this 3238\,K star), and does not clearly show an increase when using masks limited to low Land\'e factors (\geff $\leqslant 1.2$) versus high Land\'e factors (\geff $> 1.2$). The possible effect of the magnetic field on the Stokes I FWHM of the stars in this sample will be studied in detail in Donati et al. (in prep) and Cristofari et al. (in prep).

We observe that it is easier to constrain the rotation period and the other parameters of the GP for early-type stars, even if the amplitude of variation of the longitudinal magnetic field is similar in the other two subcategories. This may be due to the fact that they are brighter than later M dwarfs on average. Additionally, the decay time seems well constrained between 50 and 100 days for the early-type stars. For the mid-M dwarfs, it lies around 100 days, and for the late-M dwarfs, it seems to be longer at about 150 days, although these results are only based on three and two measurements, respectively. There is a tendency for longer decay times for a longer stellar rotation period. Clearly, the quasi-periodic GP fit converges more easily for the early-type stars than for the others: it converges for only 60\% of the mid-type stars, and although the convergence rate is similar for the late- and early-type stars, we have to use a four-parameter fit for the late-type stars (fixing the smoothing factor and the decay time), while we can use a six-parameter fit for the early-type stars with good constraints on the smoothing factor and on the decay time.

\begin{figure}
   \centering
   \includegraphics[width=\hsize]{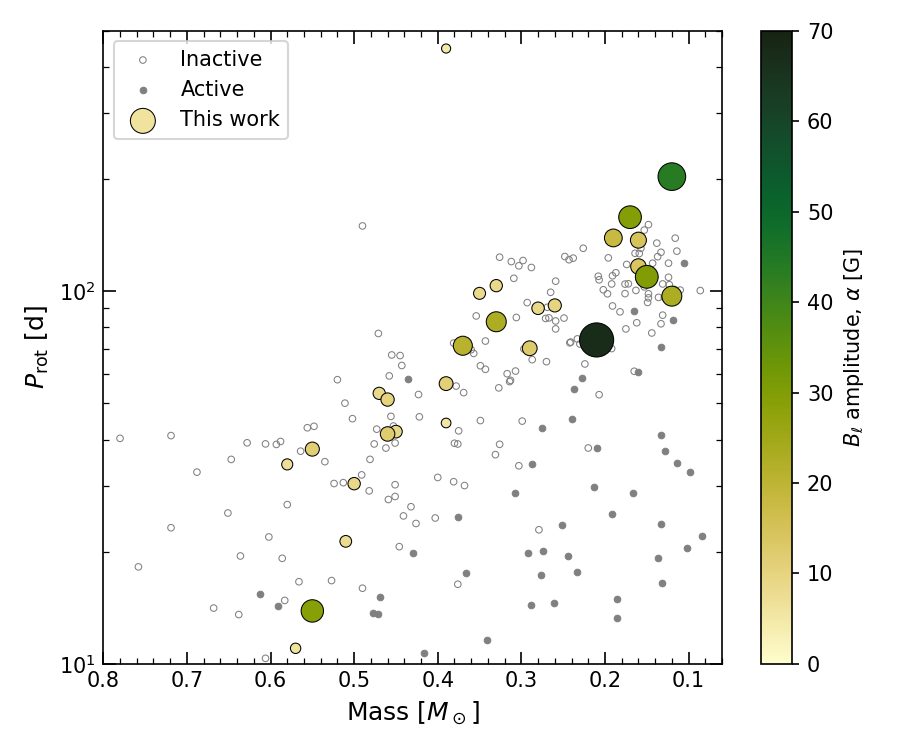}
      \caption{Rotation period vs stellar mass diagram. Stars with rotation periods determined in this work are shown as larger circles, where the symbol sizes and colors are proportional to the amplitude of variation of $B_\ell$ (see $\alpha$ values given in Tables~\ref{tab:earlygpfitparams}, \ref{tab:middlegpfitparams}, and \ref{tab:lategpfitparams}).  Rotation periods and mass determinations for inactive (white circles) and active (gray circles) stars from \citet{newton17} are shown for comparison.}
     \label{fig:protmassdiagram}
\end{figure}

Our sample is too small to infer general properties of M stars. Therefore, we compare in Fig.~\ref{fig:protmassdiagram} the sample of stars analyzed in this paper to more complete nonmagnetic studies in which active and inactive stars are well represented. We include a total of 212 M stars from \citet{newton17}, whose mass and rotation periods are similar to the range explored in this paper. Our sample of quiet stars nicely fits the sequence of inactive stars, as expected. The long rotation period of Gl\,411 appears as an exception, and may in fact not reflect the stellar rotation period, as discussed in Sec. \ref{sec:middle}.

Low-mass stars arrive rapidly rotating on the main sequence and spin down with time as angular momentum is lost through magnetized winds \citep{barnes07}.  Empirically calibrated relations between age and period can be used to estimate the ages of individual field stars. \citet{gaidos23} combined the 4\,Gyr M-dwarf gyrochrone of the open cluster M 67 \citep{dungee22} with those of younger, previously published gyrochrones \citep{curtis20} to assign ages to M dwarf host stars with $T_{\rm eff}=$3200-4200\,K.  We used the same gyrochronology to estimate the ages of 20 stars in our sample with well-measured rotation periods and $T_{\rm eff}$ in that range.  Mean values and standard deviations were calculated from distributions constructed by 10000 Monte Carlo calculations, incorporating uncertainties from $P_{\rm rot}$ (from Tables \ref{tab:middlegpfitparams} and \ref{tab:lategpfitparams}), the gyrochrones, $T_{\rm eff}$ (assumed $\pm$75K), [Fe/H] (assumed to be $\pm$0.1 dex), and variation in the initial rotation periods of the stars on the zero-age main sequence. The results are listed in Table \ref{tab:ages}: most stars are 4--10 Gyr old, and a few younger stars have ages from 0.5 to 2\,Gyr. The rotation-age relations are only valid for stars of approximately solar metallicity, but we list in the table (col. 6) a correction computed using a theory-based model and the stellar metallicity, to be added to the age given in col. 4. We report the age for members of binaries, but we emphasize that these stars may have distinct rotational histories due to tides and rapid dissipation of primordial disks \citep{fleming19, messina19}.

\begin{table*}
    \caption[]{Gyrochronological ages of the stars in our sample. Errors on $T_{\rm eff}$ and [M/H] are assumed to be 75\,K and 0.1 dex. The correction in the last column corresponds to the value to be added to the age due to the nonsolar metallicity of the star.}
    \label{tab:ages}
    \begin{center}
    \begin{tabular}{lccccc}
        \hline
        \noalign{\smallskip}
Star & $T_{\rm eff}$ & $P_{\rm rot}$ & Age &  [M/H] & Metallicity correction \\
& K & d & Gyr & dex & Gyr \\
        \noalign{\smallskip}
        \hline
        \noalign{\smallskip}
Gl\,846 & 3833 & $11.01 \pm 0.20$ & $<0.53$ & 0.07 & 0.1 \\
Gl\,410 & 3842 & $13.87 \pm 0.08$ & $0.89 \pm 0.12$ & 0.05 & 0.1 \\
Gl\,382 & 3644 & $21.32 \pm 0.04$ & $1.9 \pm 0.6$ & 0.15 & 0.2 \\
Gl\,514 & 3699 & $30.45 \pm 0.14$ & $3.8 \pm 0.6$ & $-0.07$ & $-0.1$ \\
GJ\,1289 & 3238 & $74.0 \pm 1.4$ & $3.8 \pm 1.2$ & 0.05 & 0.0 \\
Gl\,849 & 3502 & $41.4 \pm 0.4$ & $4.2 \pm 1.0$ & 0.35 & 0.4 \\
Gl\,205 & 3771 & $34.3 \pm 0.4$ & $5.2 \pm 0.7$ & 0.43 & 0.6 \\
PM J09553-2715 & 3366 & $70.5 \pm 3.8$ & $5.2 \pm 2.0$ & $-0.03$ & $-0.0$ \\
Gl\,880 & 3702 & $37.7 \pm 0.7$ & $5.5 \pm 0.8$ & 0.26 & 0.4 \\
Gl\,169.1A & 3307 & $91.9 \pm 3.4$ & $5.5 \pm 2.4$ & 0.13 & 0.1 \\
Gl\,687 & 3475 & $56.5 \pm 0.9$ & $5.9 \pm 1.8$ & 0.01 & 0.0 \\
Gl\,15A & 3611 & $44.3 \pm 0.2$ & $6.0 \pm 1.2$ & $-0.33$ & $-0.5$ \\
GJ\,3378 & 3326 & $92.1 \pm 4.7$ & $6.0 \pm 2.7$ & $-0.05$ & $-0.0$ \\
Gl\,48 & 3529 & $51.2 \pm 1.4$ & $6.1 \pm 1.6$ & 0.08 & 0.1 \\
Gl\,15B & 3272 & $116.5 \pm 5.6$ & $6.1 \pm 3.0$ & $-0.42$ & $-0.2$ \\
Gl\,876 & 3366 & $82.8 \pm 1.4$ & $6.2 \pm 2.7$ & 0.15 & 0.1 \\
Gl\,752A & 3558 & $53.2 \pm 4.2$ & $7.1 \pm 1.9$ & 0.11 & 0.2 \\
Gl\,699 & 3311 & $137.1 \pm 5.4$ & $9.4 \pm 5.0$ & $-0.37$ & $-0.2$ \\
Gl\,251 & 3420 & $98.7 \pm 8.2$ & $10.8 \pm 4.7$ & $-0.01$ & $-0.0$ \\
Gl\,725A & 3470 & $103.5 \pm 4.8$ & $14.8 \pm 5.4$ & $-0.26$ & $-0.3$ \\
        \hline
    \end{tabular}
    \end{center}
  \end{table*}

Our study only exploited the longitudinal magnetic field measured from the spectropolarimetric data of the SPIRou SLS. The same data can be used to achieve much more, for example, to investigate the magnetic topology of these stars and its evolution using ZDI maps.

\begin{acknowledgements}

Based on observations obtained at the Canada-France-Hawaii Telescope (CFHT) which is operated from the summit of Maunakea by the National Research Council of Canada, the \emph{Institut National des Sciences de l'Univers} of the \emph{Centre National de la Recherche Scientifique} of France, and the University of Hawaii. The observations at the Canada-France-Hawaii Telescope were performed with care and respect from the summit of Maunakea which is a significant cultural and historic site. \\

This work has made use of data from the European Space Agency (ESA) mission {\it Gaia} (\url{https://www.cosmos.esa.int/gaia}), processed by the {\it Gaia} Data Processing and Analysis Consortium (DPAC, \url{https://www.cosmos.esa.int/web/gaia/dpac/consortium}). Funding for the DPAC has been provided by national institutions, in particular, the institutions participating in the {\it Gaia} Multilateral Agreement.\\

This research has also made intensive use of the SIMBAD database and of the VizieR catalog access tool, operated at CDS, Strasbourg, France and of NASA's Astrophysics Data System (ADS). \\

We made an extensive use of the DACE (Data Analysis Center for Exoplanets) software from the University of Geneva \citep{diaz14, delisle16}. \\

This work has made use of the VALD database, operated at Uppsala University, the Institute of Astronomy RAS in Moscow, and the University of Vienna. \\

X.D., A.C., P.C.Z., E.M. and S.B. and more generally most of the French authors of this paper acknowledge funding from the French National Research Agency (ANR) under contract number ANR\-18\-CE31\-0019 (SPlaSH). This work is supported by the ANR in the framework of the \emph{Investissements d'Avenir} program (ANR-15-IDEX-02), through the funding of the ``Origin of Life'' project of the Grenoble-Alpes University. \\
 
E.M. acknowledges funding from the \emph{Funda\c{c}\~{a}o de Amparo \`{a} Pesquisa do Estado de Minas Gerais} (FAPEMIG) under the project number APQ-02493-22. \\

B.Z. acknowledges funding from the \emph{Programa de Internacionaliza\c{c}\~{a}o da Coordena\c{c}\~{a}o de Aperfei\c{c}oamento de Pessoal de Nível Superior} (CAPES-PrInt  \#88887.683070/2022-00) and FAPEMIG (APQ-01033-22). \\

J.-F.D. acknowledges funding from the European Research Council (ERC) under the H2020 research \& innovation programme (grant agreement \#740651 NewWorlds). \\

N.J.C., E.A. and R.D. wish to thank the Natural Sciences and Engineering Research Council of Canada and the \emph{Fonds Qu\'eb\'ecois de Recherche - Nature et Technologies}, the \emph{Observatoire du Mont-M\'egantic} and the Institute for Research on Exoplanets and acknowledge funding from \emph{D\'eveloppement Economique Canada, Quebec's Minist\`ere de l'Education et de l'Innovation}, the Trottier Family Foundation and the Canadian Space Agency. \\

This research made use of the hosting service \texttt{github} and, among others, of the following software tools: \verb|matplotlib| \citep{hunter07}; \verb|NumPy| \citep{harris20}; \verb|SciPy| \citep{scipy20}; \verb|Astropy| \citep{astropy13,astropy18};  \verb|emcee| \citep{foreman13};  \verb|corner| \citep{foreman16}; \verb|george| \citep{ambikasaran15}; \verb|barrycorpy| \citep{wright14}.

\end{acknowledgements}

%
%

\bibliographystyle{aa}
\bibliography{45839corr}

\appendix

\section{Quasi-periodic Gaussian process regression for stars with a measured rotation period}
\label{sec:GPfit}
\begin{figure}
     \centering
       \includegraphics[width=1.0\hsize]{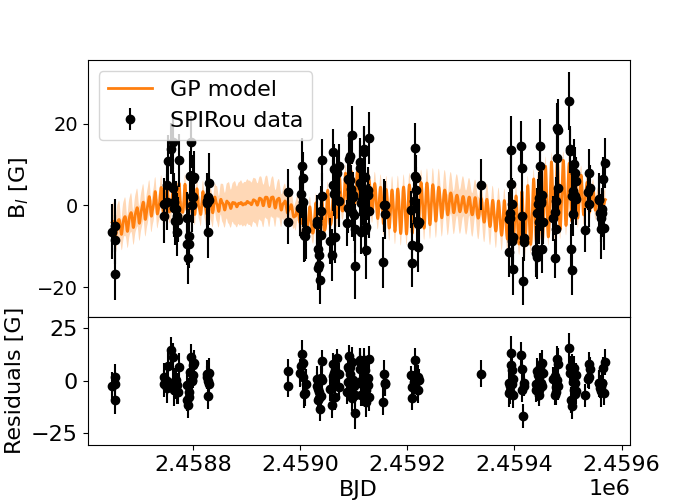}
      \caption{GP analysis of the SPIRou $B_\ell$ data of Gl\,846. The top panel shows the observed $B_\ell$ data (black points) and the orange line shows the best-fit quasi-periodic GP model. The bottom panel shows the residuals.}
        \label{fig:GL846_blong_GP}
\end{figure}

\begin{figure}
     \centering
       \includegraphics[width=1.0\hsize]{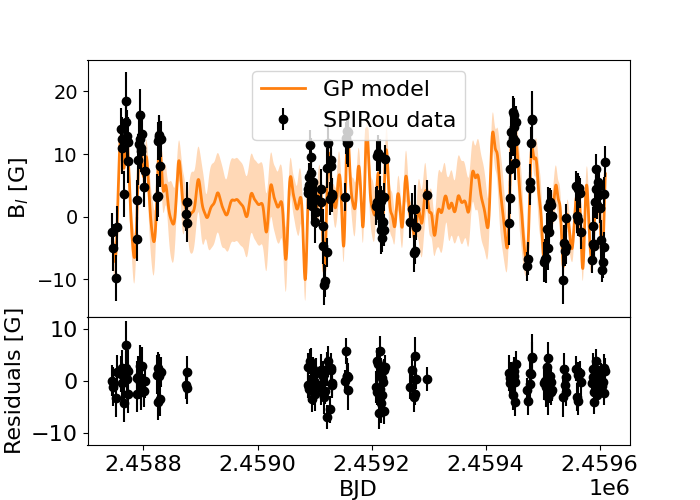}
      \caption{Same for Gl\,205.}
        \label{fig:GL205_blong_GP}
\end{figure}

\begin{figure}
     \centering
       \includegraphics[width=1.0\hsize]{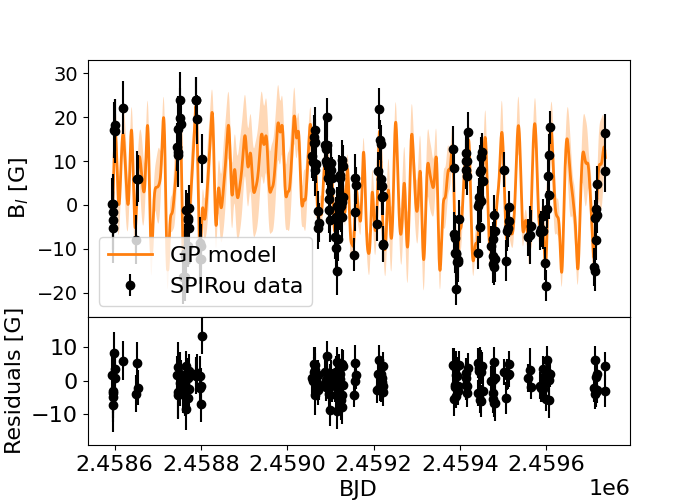}
      \caption{Same for Gl\,880.}
        \label{fig:GL880_blong_GP}
\end{figure}

\begin{figure}
     \centering
       \includegraphics[width=1.0\hsize]{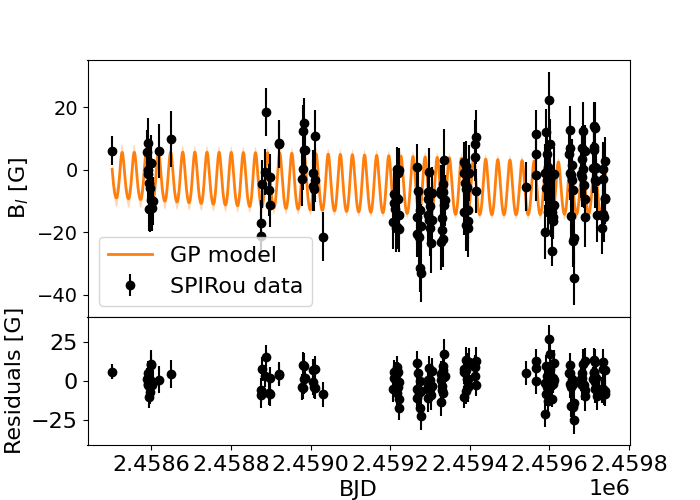}
      \caption{Same for Gl\,514.}
        \label{fig:GL514_blong_GP}
\end{figure}

\begin{figure}
     \centering
       \includegraphics[width=1.0\hsize]{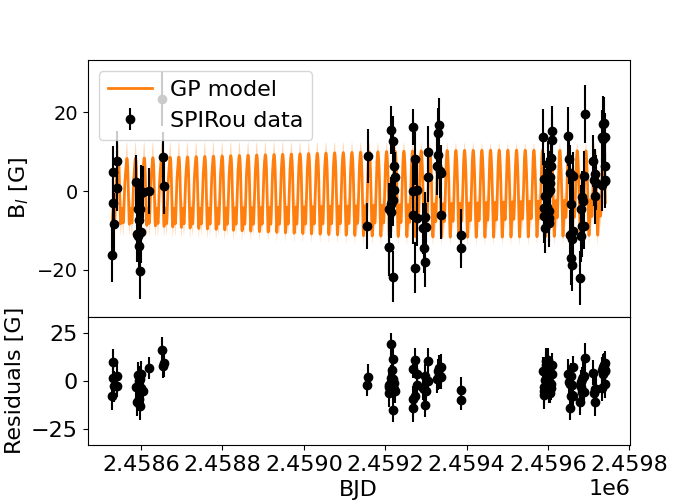}
      \caption{Same for Gl\,382.}
        \label{fig:GL382_blong_GP}
\end{figure}

\begin{figure}
     \centering
       \includegraphics[width=1.0\hsize]{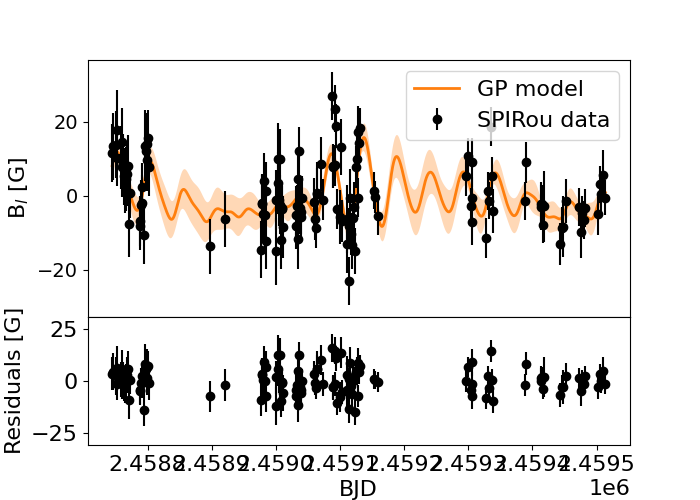}
      \caption{Same for Gl\,752A.}
        \label{fig:GL752A_blong_GP}
\end{figure}

\begin{figure}
     \centering
       \includegraphics[width=1.0\hsize]{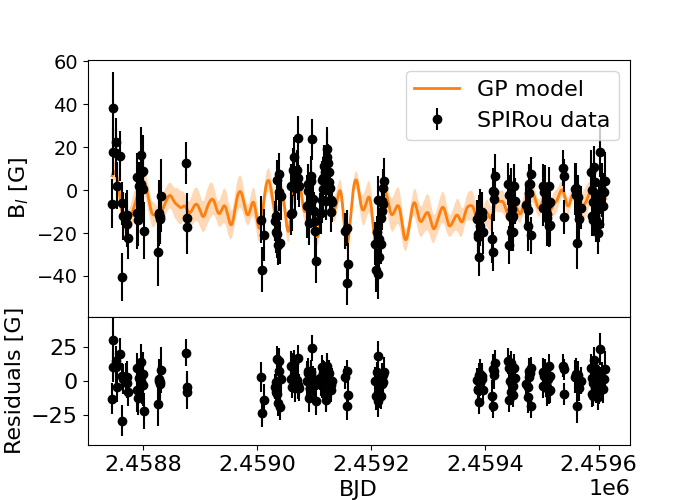}
      \caption{Same for Gl\,48.}
        \label{fig:GL48_blong_GP}
\end{figure}

\begin{figure}
     \centering
       \includegraphics[width=1.0\hsize]{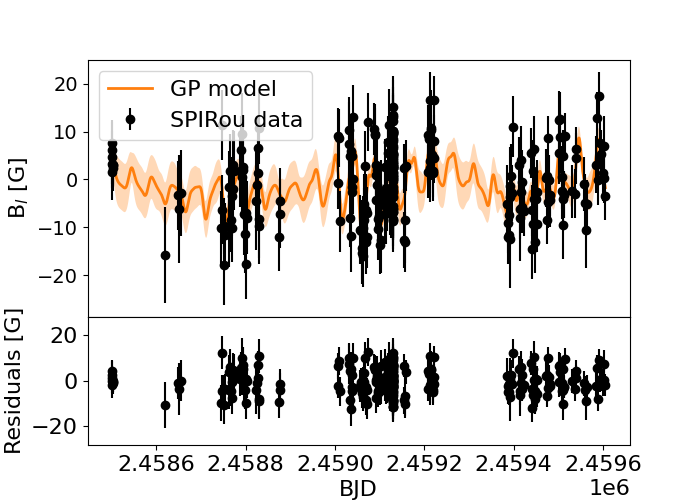}
      \caption{Same for Gl\,15A.}
        \label{fig:GL15A_blong_GP}
\end{figure}

\begin{figure}
     \centering
       \includegraphics[width=1.0\hsize]{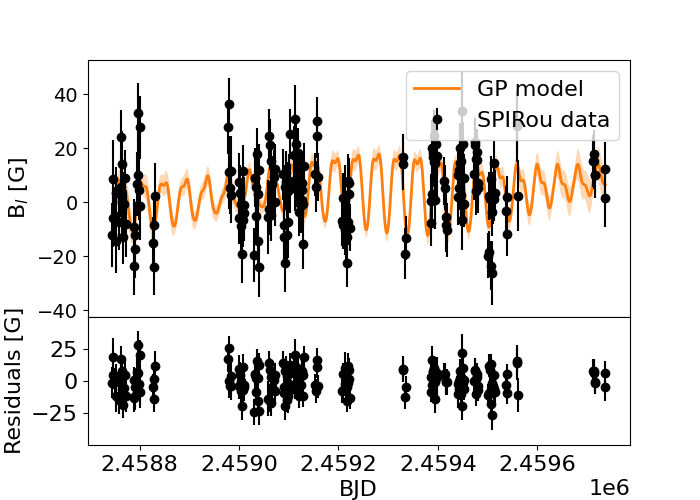}
      \caption{Same for Gl\,849.}
        \label{fig:GL849_blong_GP}
\end{figure}

\begin{figure}
     \centering
       \includegraphics[width=1.0\hsize]{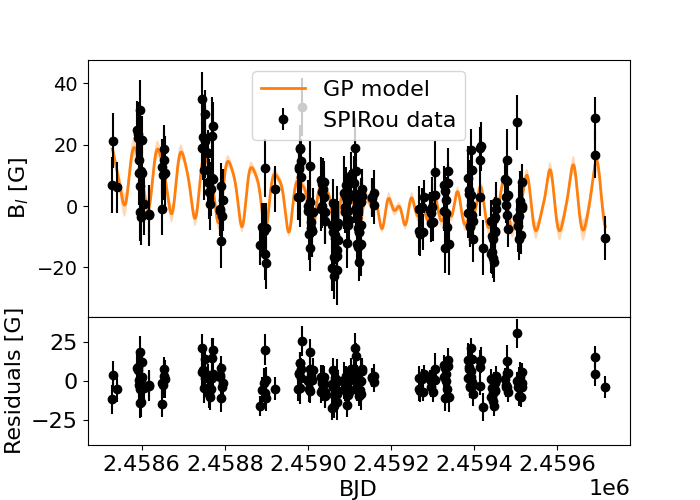}
      \caption{Same for Gl\,687.}
        \label{fig:GL687_blong_GP}
\end{figure}

\begin{figure}
     \centering
       \includegraphics[width=1.0\hsize]{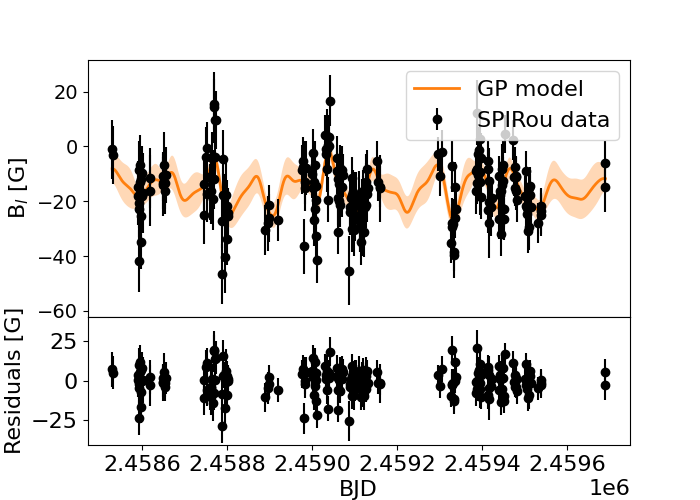}
      \caption{Same for Gl\,725A (five-parameter fit).}
        \label{fig:GL725A_blong_GP}
\end{figure}

\begin{figure}
     \centering
       \includegraphics[width=1.0\hsize]{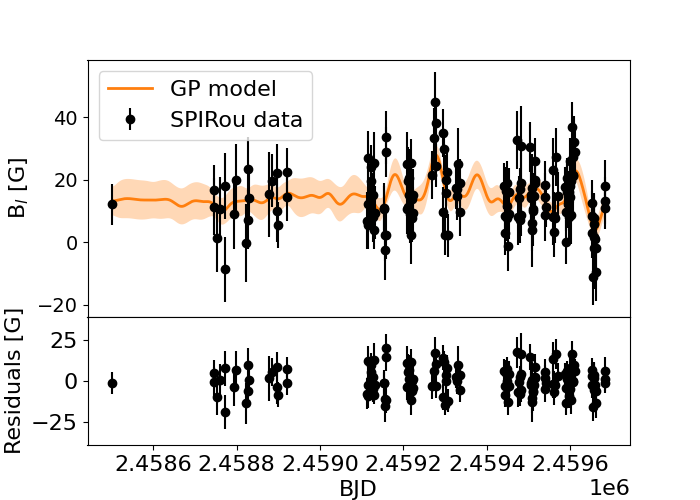}
      \caption{Same for Gl\,251 (five-parameter fit).}
        \label{fig:GL251_blong_GP}
\end{figure}

\begin{figure}
     \centering
       \includegraphics[width=1.0\hsize]{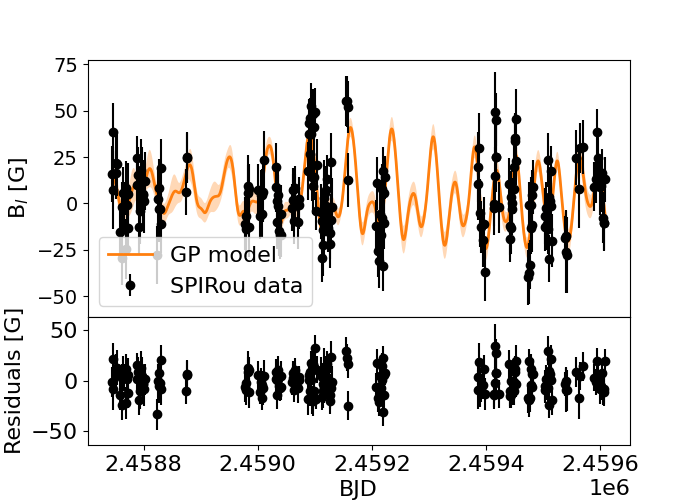}
      \caption{Same for GJ\,4333.}
        \label{fig:GJ4333_blong_GP}
\end{figure}

\begin{figure}
     \centering
       \includegraphics[width=1.0\hsize]{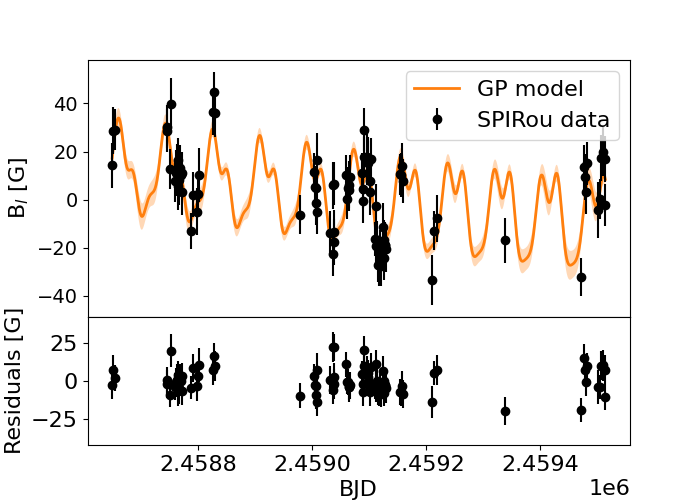}
      \caption{Same for Gl\,876.}
        \label{fig:GL876_blong_GP}
\end{figure}

\begin{figure}
     \centering
       \includegraphics[width=1.0\hsize]{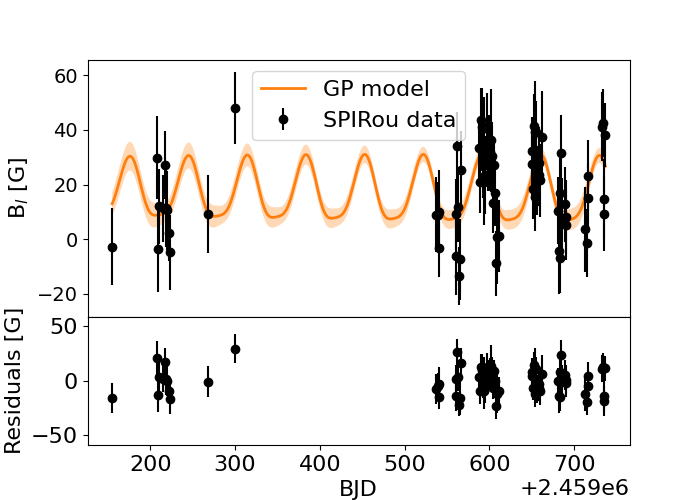}
      \caption{Same for PM\,J09553-2715.}
        \label{fig:PM_J09553-2715_blong_GP}
\end{figure}

\begin{figure}
     \centering
       \includegraphics[width=1.0\hsize]{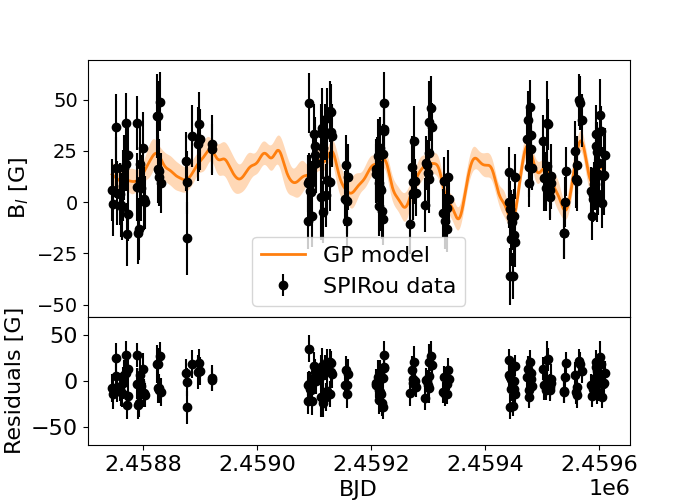}
      \caption{Same for GJ\,3378.}
        \label{fig:GJ3378_blong_GP}
\end{figure}

\begin{figure}
     \centering
       \includegraphics[width=1.0\hsize]{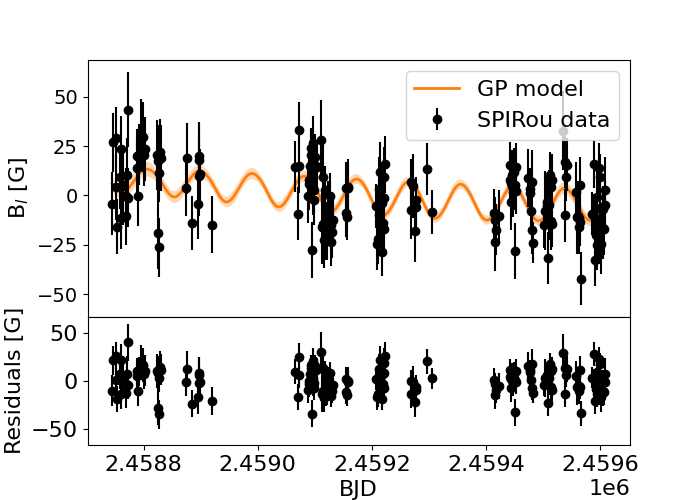}
      \caption{Same for Gl\,169.1A.}
        \label{fig:GL169.1A_blong_GP}
\end{figure}

\begin{figure}
     \centering
       \includegraphics[width=1.0\hsize]{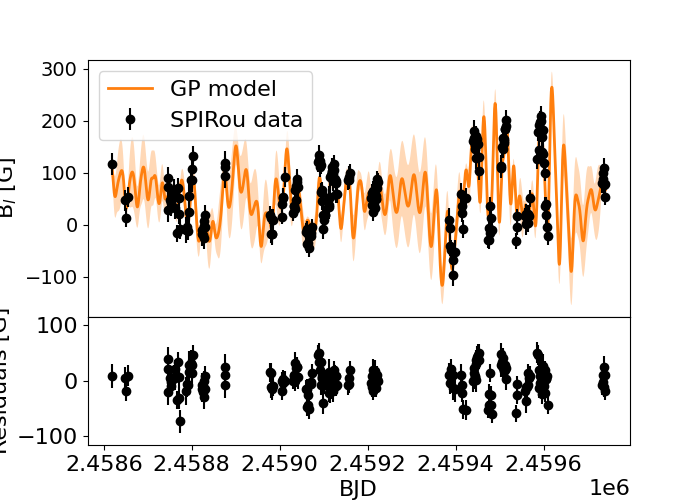}
      \caption{Same for GJ\,1289.}
        \label{fig:GJ1289_blong_GP}
\end{figure}

\begin{figure}
     \centering
       \includegraphics[width=1.0\hsize]{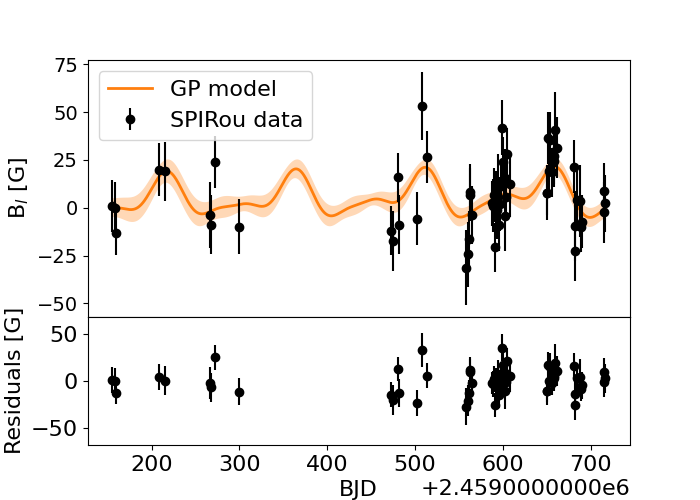}
      \caption{Same for GJ\,1103 (five-parameter fit).}
        \label{fig:GJ1103_blong_GP}
\end{figure}

\begin{figure}
     \centering
       \includegraphics[width=1.0\hsize]{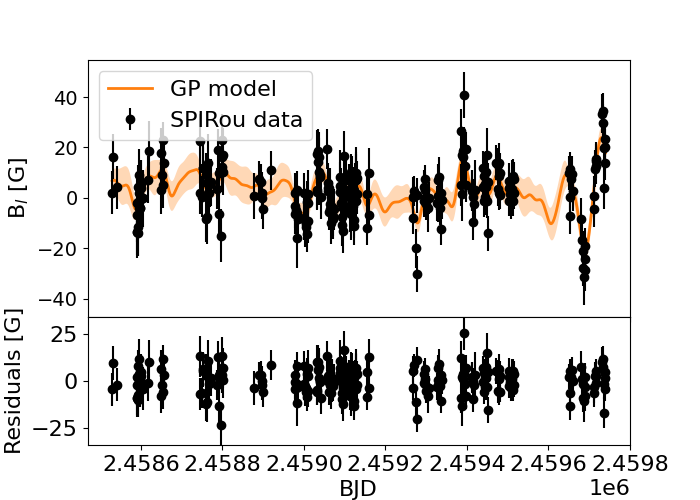}
      \caption{Same for Gl\,699 (five-parameter fit).}
        \label{fig:GL699_blong_GP}
\end{figure}

\begin{figure}
     \centering
       \includegraphics[width=1.0\hsize]{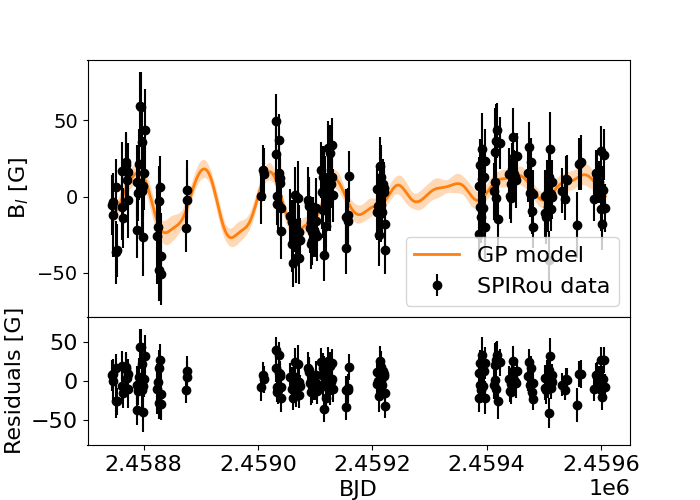}
      \caption{Same for Gl\,15B (five-parameter fit).}
        \label{fig:GL15B_blong_GP}
\end{figure}

\begin{figure}
     \centering
       \includegraphics[width=1.0\hsize]{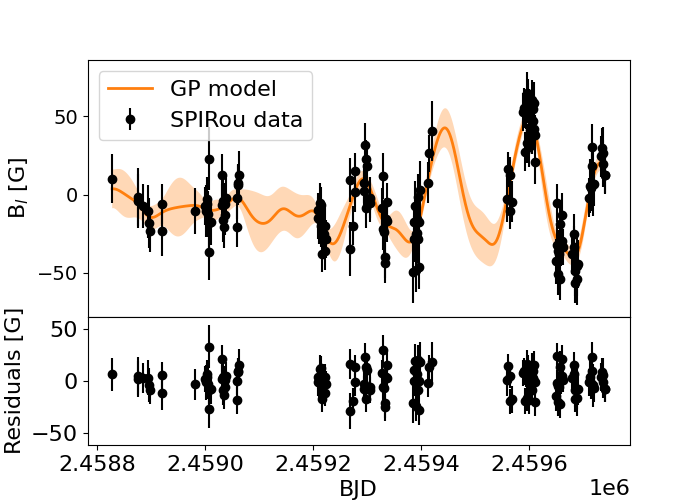}
      \caption{Same for GJ\,1151 (five-parameter fit).}
        \label{fig:GJ1151_blong_GP}
\end{figure}

\begin{figure}
     \centering
       \includegraphics[width=1.0\hsize]{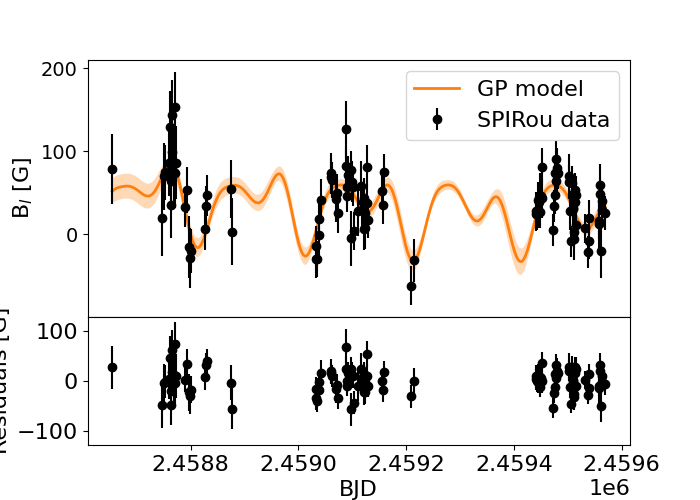}
      \caption{Same for GJ\,1286 (five-parameter fit).}
        \label{fig:GJ1286_blong_GP}
\end{figure}

\begin{figure}
     \centering
       \includegraphics[width=1.0\hsize]{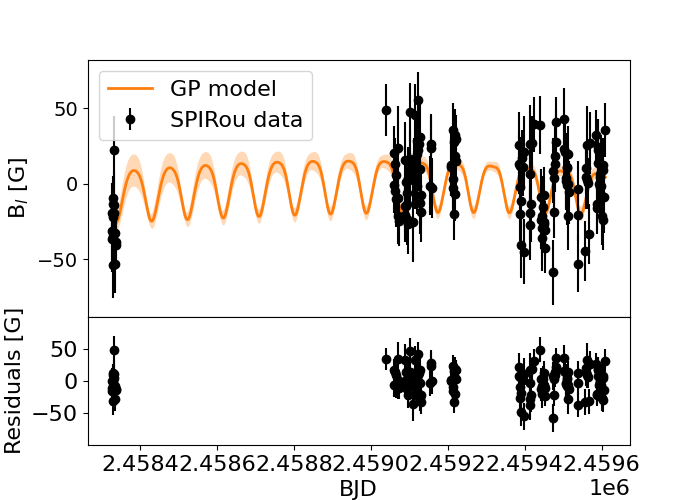}
      \caption{Same for GJ\,1002.}
        \label{fig:GJ1002_blong_GP}
\end{figure}

We display here the temporal evolution of the longitudinal magnetic field and its GP fit leading to the measurement of the stellar rotation period for 25 stars of our sample (Figs.~\ref{fig:GL846_blong_GP} to \ref{fig:GJ1002_blong_GP}). GP fits are generally six-parameter fits, except when stated in the caption.

\clearpage

\clearpage
\section{Quasi-periodic Gaussian process regression for Gl 411}
\label{sec:GPfit411}
\begin{figure*}
     \centering
       \includegraphics[width=1.0\hsize]{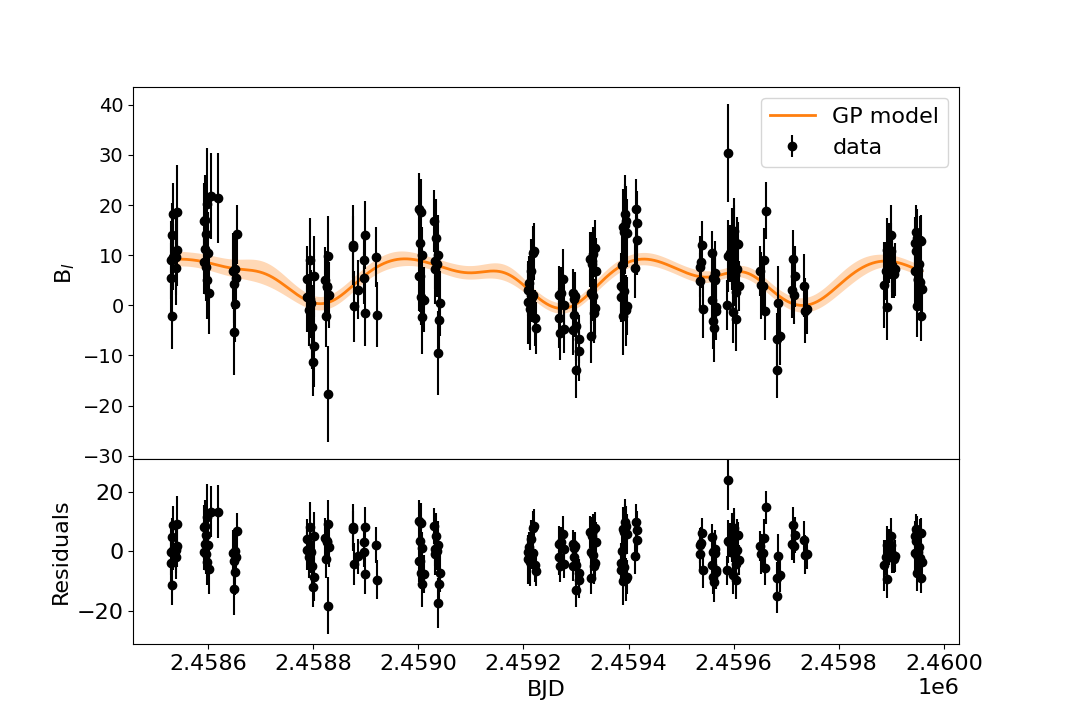}
      \caption{GP analysis of the SPIRou $B_\ell$ data of Gl\,411. The top panel shows the observed $B_\ell$ data (black points) and the orange line shows the best-fit quasi-periodic GP model. The bottom panel shows the residuals of the six-parameter GP fit. }
        \label{fig:GL411_blong_GP}
\end{figure*}

In this appendix, we study the particular case of the very long stellar rotation period measured for Gl\,411 (460 days). We display the temporal evolution of the longitudinal magnetic field and its GP fit (Fig.~\ref{fig:GL411_blong_GP}). We then display the associated corner plot showing the posterior distributions of each parameter (Fig.~\ref{fig:GL411_blong_CP}).

\begin{figure*}
     \centering
       \includegraphics[width=1.0\hsize]{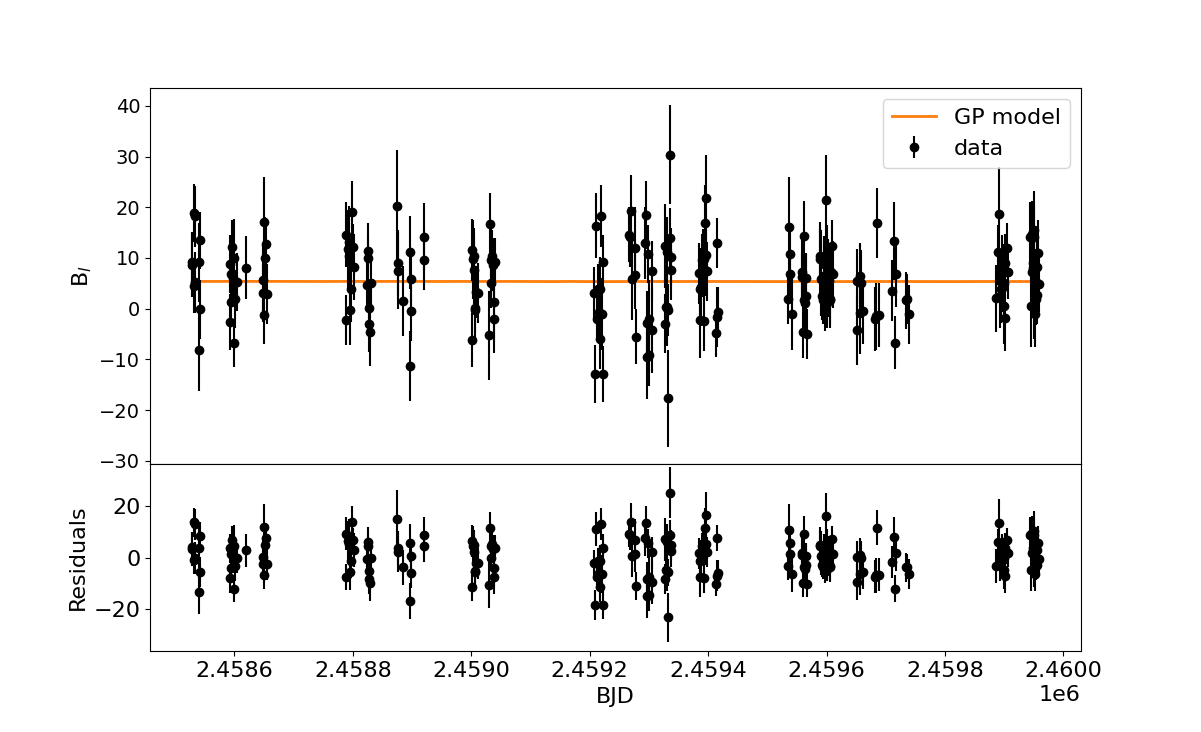}
      \caption{GP analysis of reshuffled $B_\ell$ data at the same observation dates as Gl\,411.}
        \label{fig:GL411_blong_shuffle_GP}
\end{figure*}

\begin{figure*}
     \centering
       \includegraphics[width=1.0\hsize]{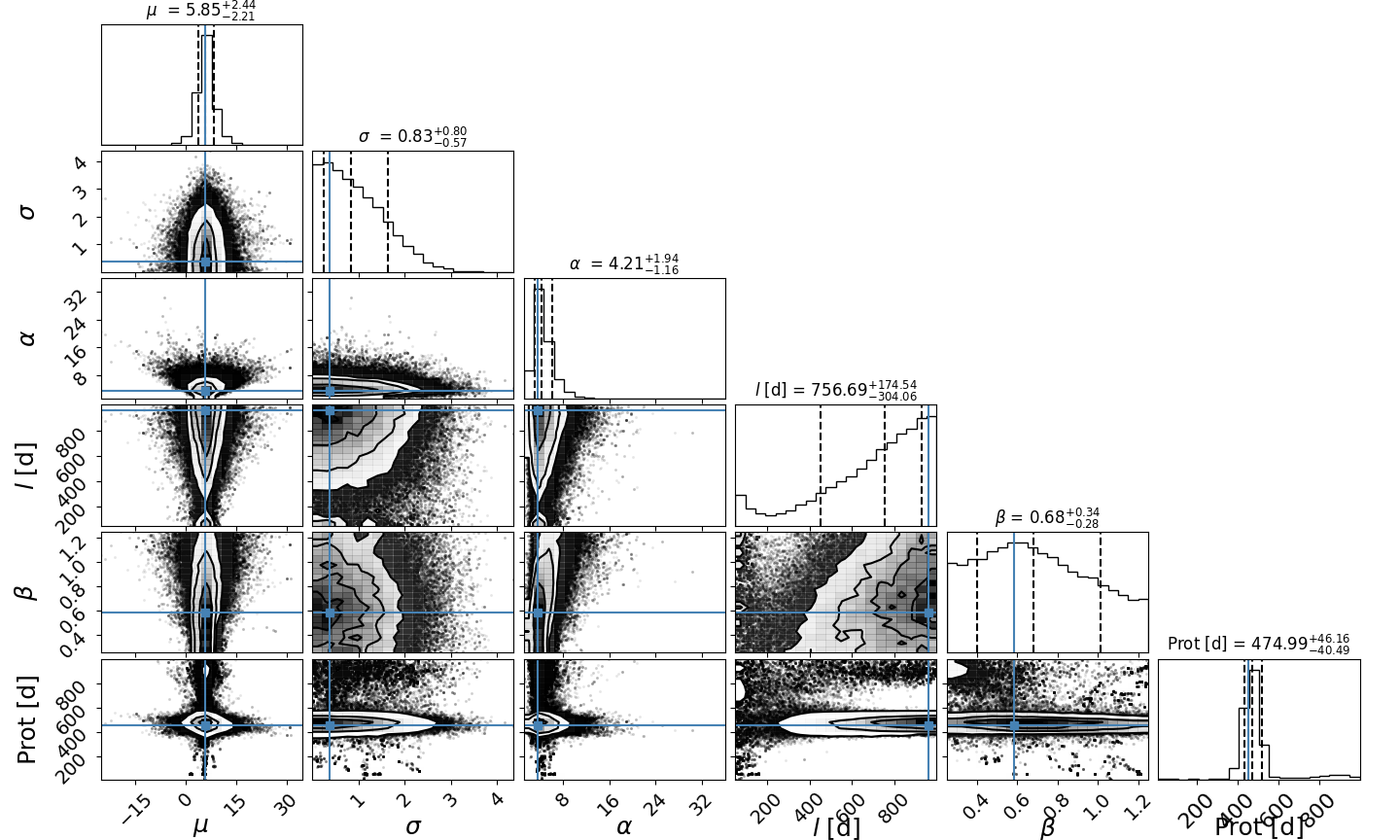}
      \caption{Corner plot of the posterior distribution of parameters of the GP fit for Gl\,411.}
        \label{fig:GL411_blong_CP}
\end{figure*}

\begin{figure*}
     \centering
       \includegraphics[width=1.0\hsize]{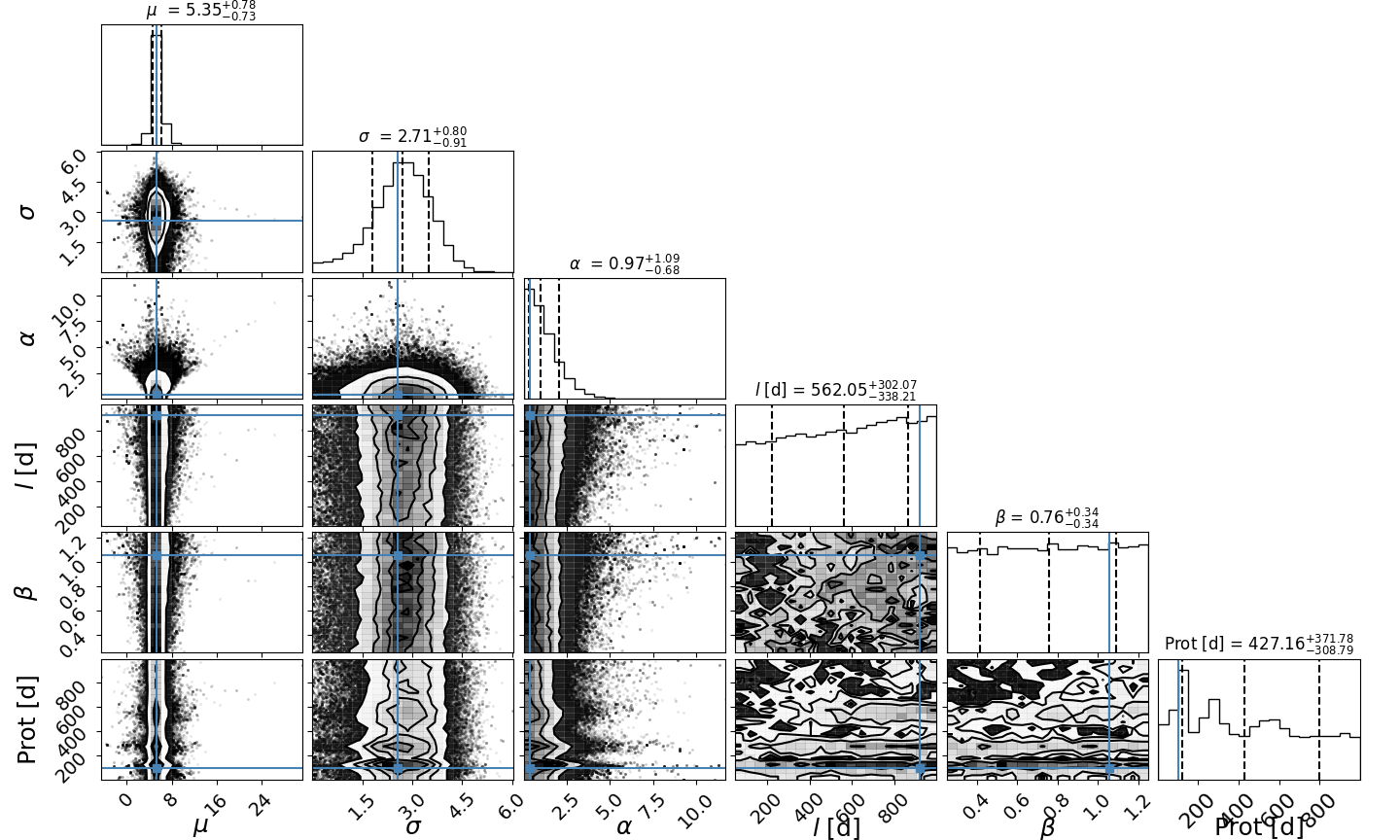}
      \caption{Corner plot of the posterior distribution of parameters of the GP fit for the reshuffled data at the observation dates of Gl\,411.}
        \label{fig:GL411_blong_shuffle_CP}
\end{figure*}

At the suggestion of the referee, we compared the GP fit to a similar GP fit to reshuffled data. We kept the same observation dates, but attributed the $B_\ell$ measurements and their associated uncertainty randomly to a given date. The GP fit is now flat and shows that the detected periodicity in the actually measured data is real (Fig.~\ref{fig:GL411_blong_shuffle_GP}). We also display the associated corner plot for comparison. No periodic behavior can be detected at all (Fig.~\ref{fig:GL411_blong_shuffle_CP}).

\clearpage
\section{Stokes V profiles of stars without a detected rotation period}
\label{sec:stokesV}
\begin{figure}
   \centering
   \includegraphics[width=0.9\hsize]{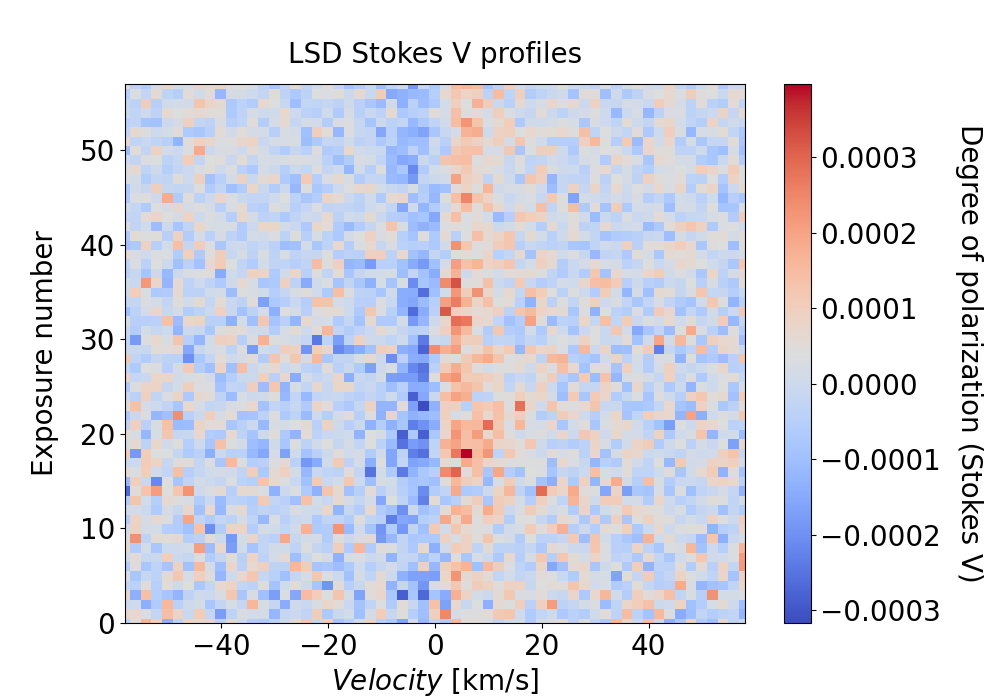}
      \caption{Temporal variation in Stokes $V$ profiles of the SPIRou polarimetric series of Gl\,338B.}
   \label{fig:gl338b}
\end{figure}

\begin{figure}
   \centering
   \includegraphics[width=0.9\hsize]{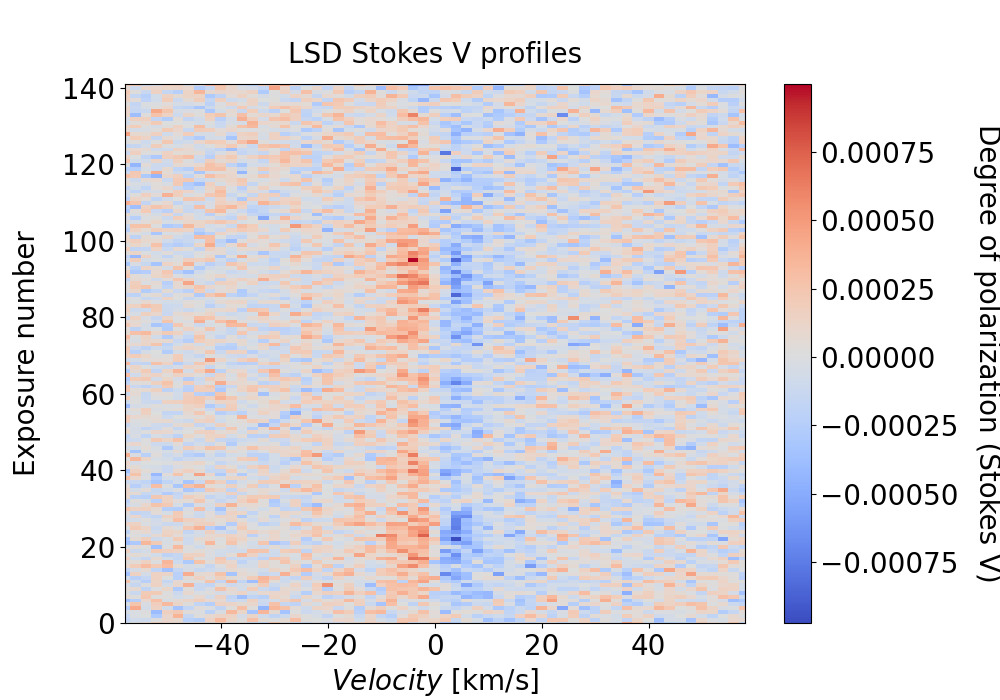}
      \caption{Same for Gl\,617B.}
    \label{fig:gl617b}
\end{figure}

\begin{figure}
   \centering
   \includegraphics[width=0.9\hsize]{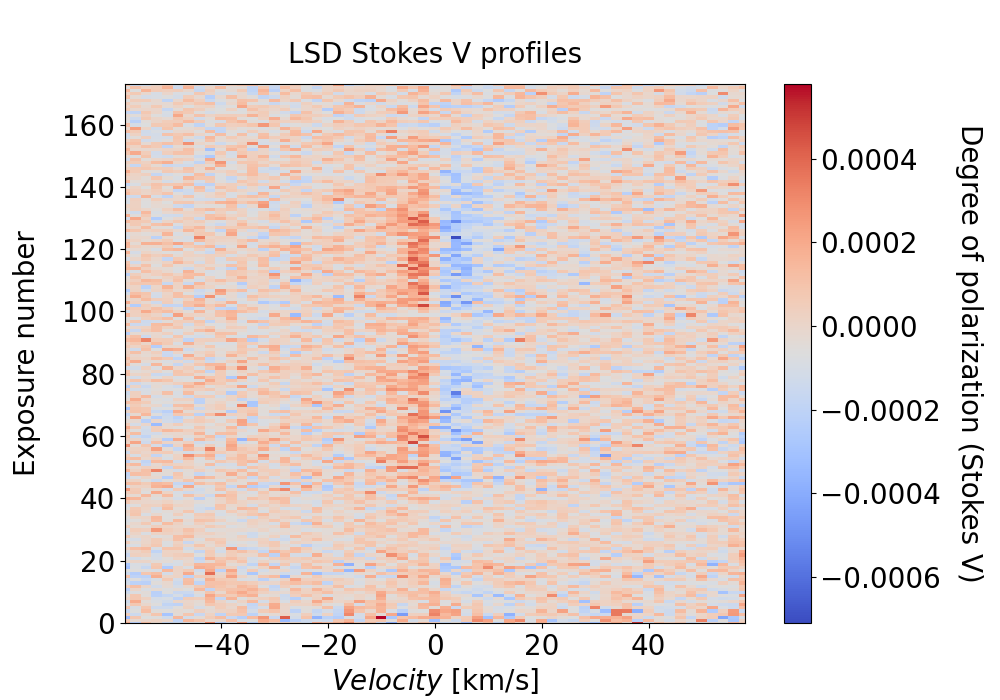}
      \caption{Same for Gl\,412A.}
    \label{fig:gl412a}
\end{figure}

\begin{figure}
   \centering
   \includegraphics[width=0.9\hsize]{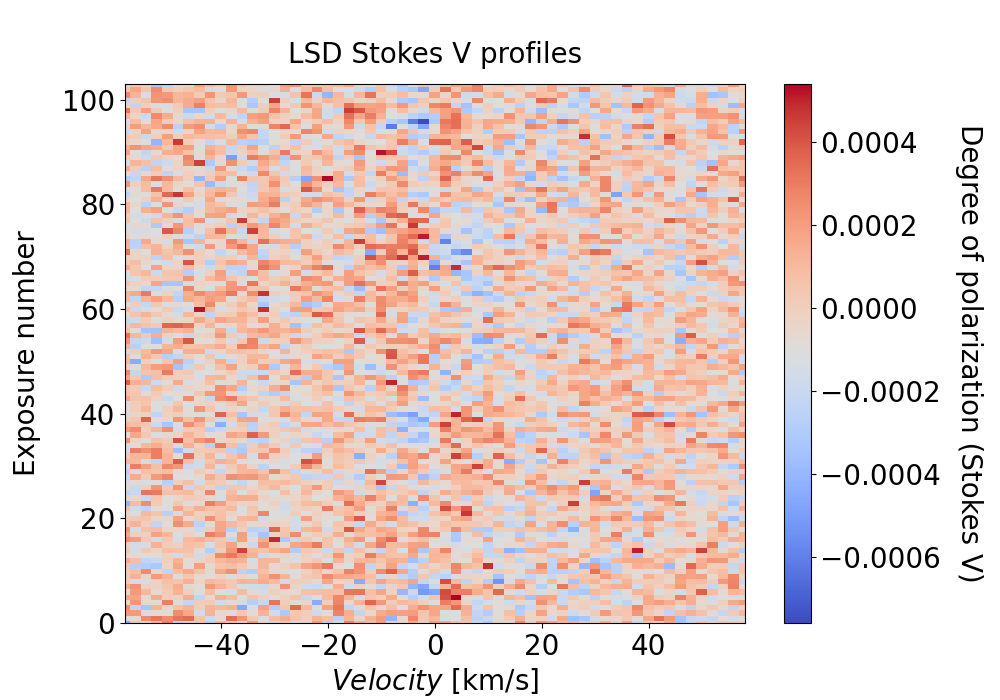}
      \caption{Same for Gl\,480.}
    \label{fig:gl480}
\end{figure}

\begin{figure}
   \centering
   \includegraphics[width=0.9\hsize]{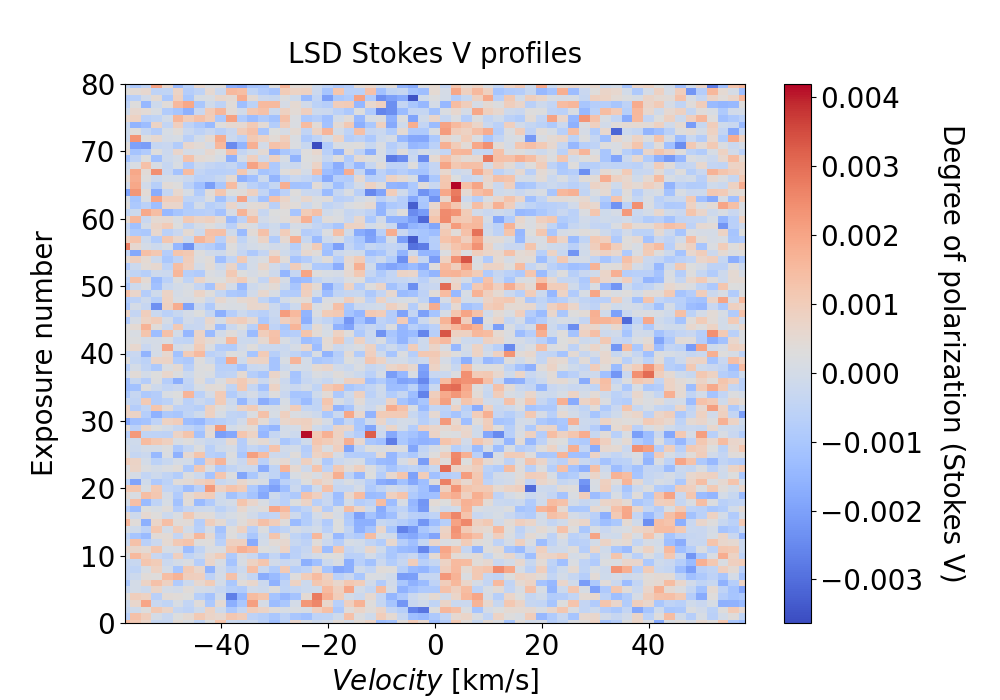}
      \caption{Same for Gl\,436.}
    \label{fig:gl436}
\end{figure}

\begin{figure}
   \centering
   \includegraphics[width=0.9\hsize]{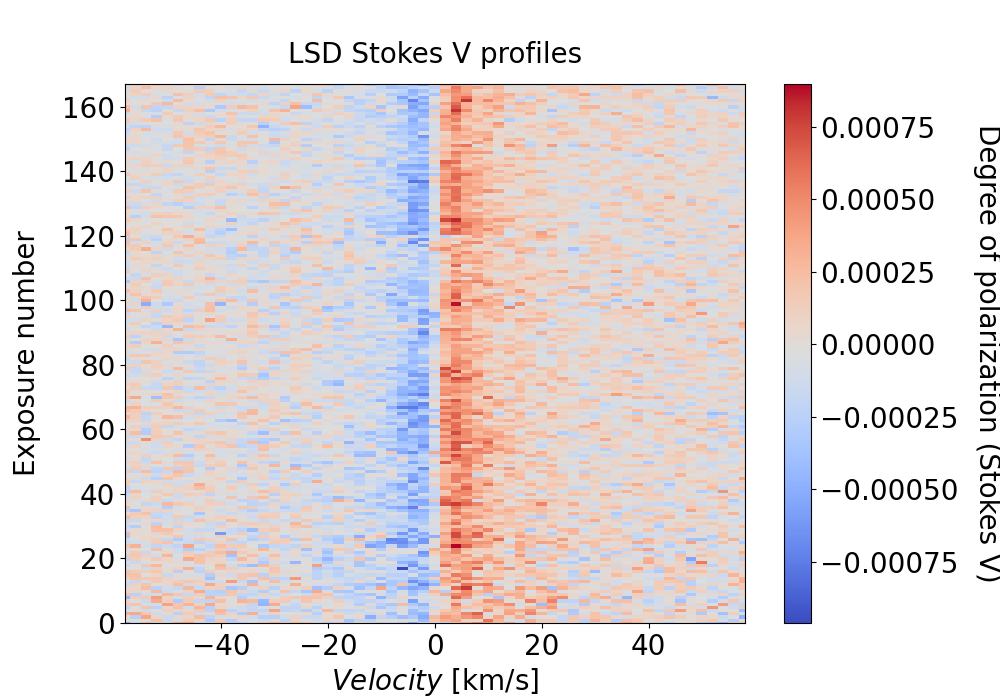}
      \caption{Same for Gl\,408.}
    \label{fig:gl408}
\end{figure}

\begin{figure}
   \centering
   \includegraphics[width=0.9\hsize]{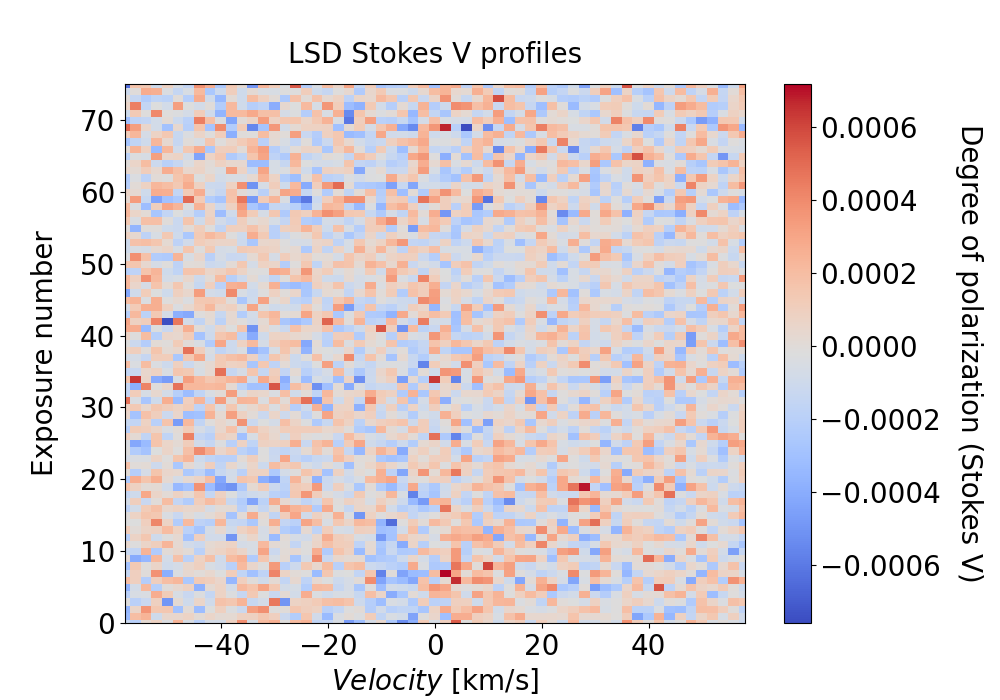}
      \caption{Same for Gl\,317.}
    \label{fig:gl317}
\end{figure}

\begin{figure}
   \centering
   \includegraphics[width=0.9\hsize]{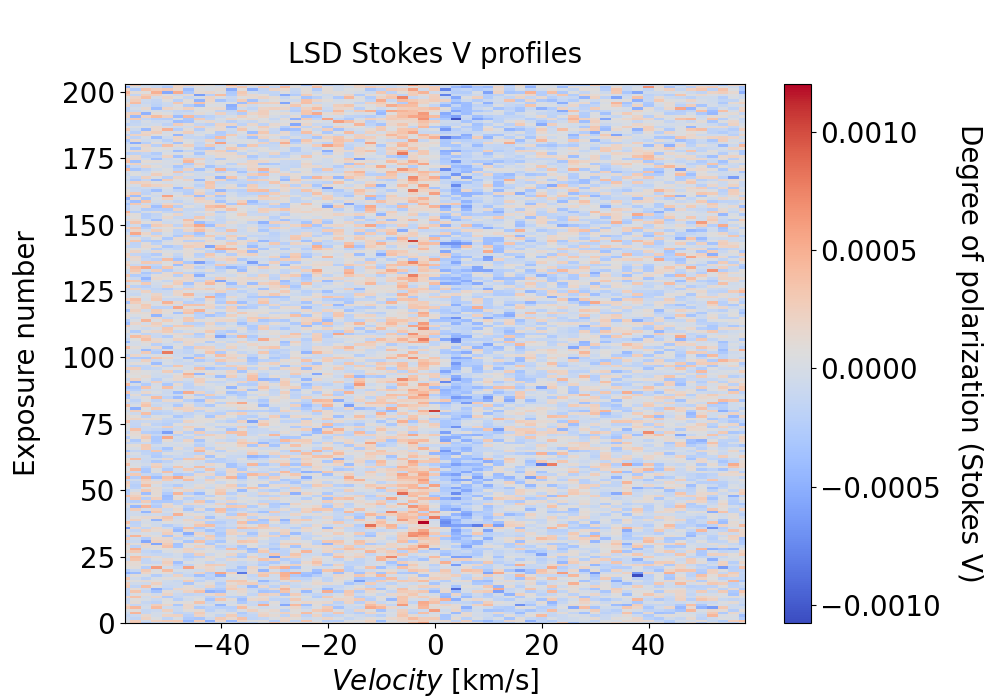}
      \caption{Same for GJ\,4063.}
    \label{fig:gl4063}
\end{figure}

\begin{figure}
   \centering
   \includegraphics[width=0.9\hsize]{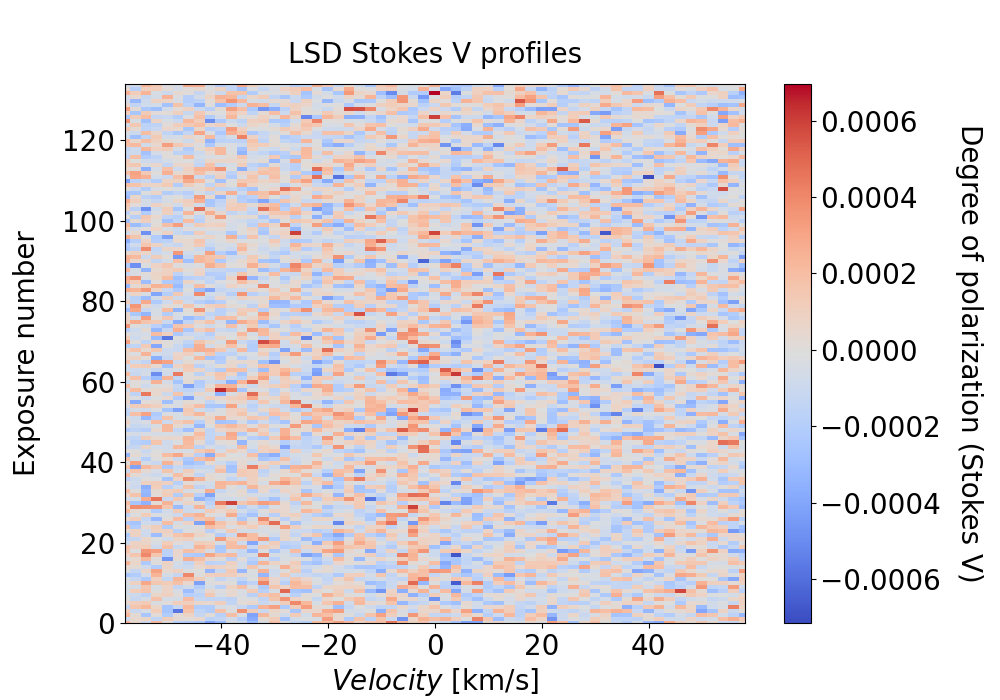}
      \caption{Same for GJ\,1012.}
    \label{fig:gl1012}
\end{figure}

\begin{figure}
   \centering
   \includegraphics[width=0.9\hsize]{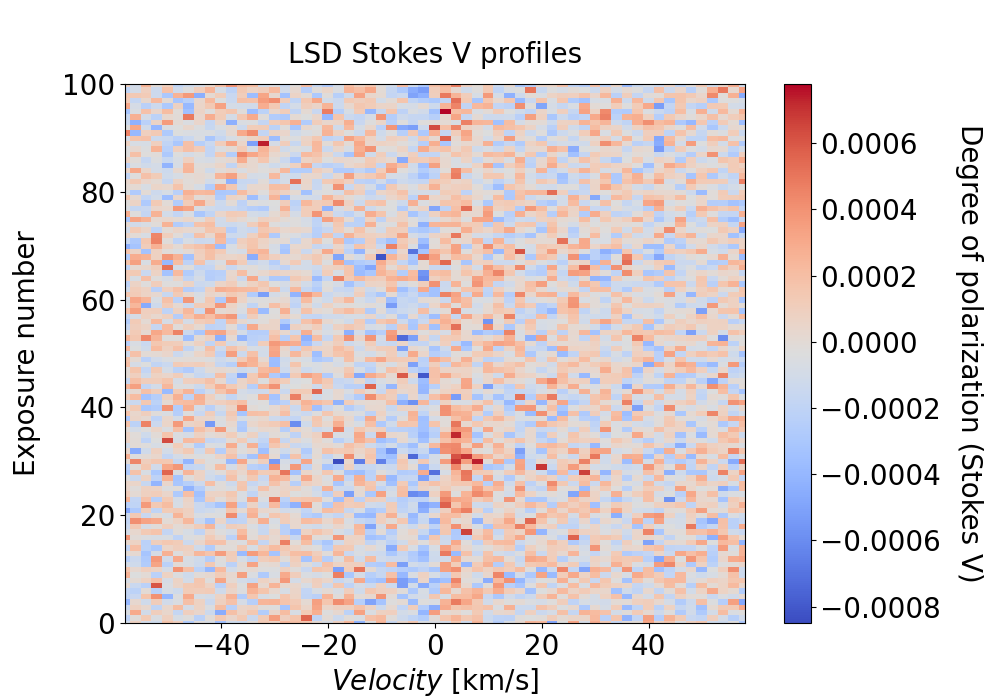}
      \caption{Same for GJ\,1148.}
    \label{fig:gl1148}
\end{figure}

\begin{figure}
   \centering
   \includegraphics[width=0.9\hsize]{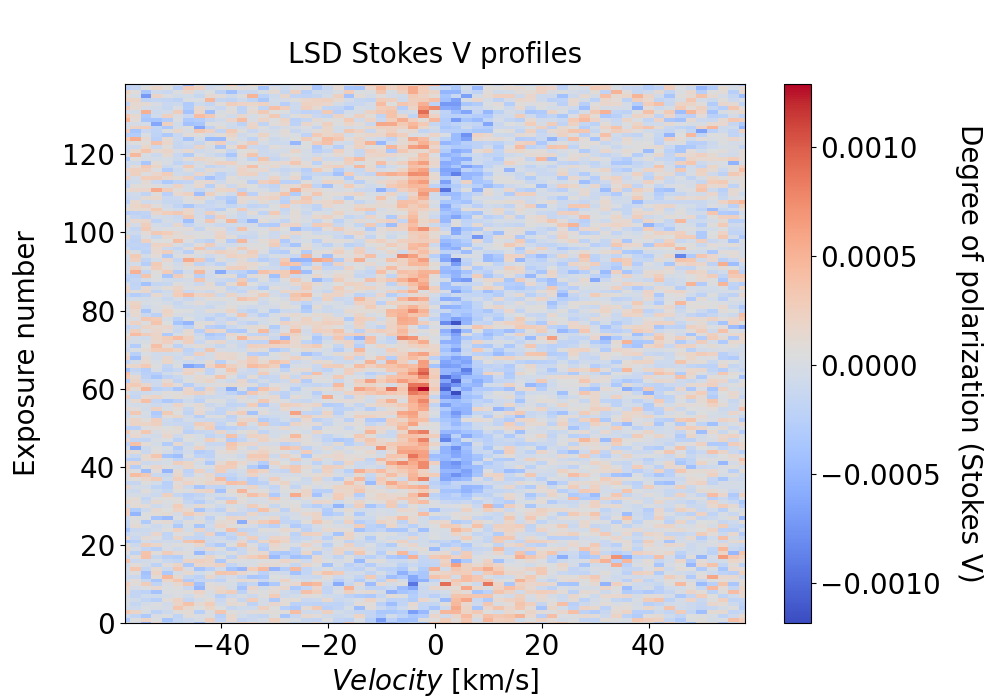}
      \caption{Same for PM\,J08402+3127.}
    \label{fig:pmJ08402+3127}
\end{figure}

\begin{figure}
   \centering
   \includegraphics[width=0.9\hsize]{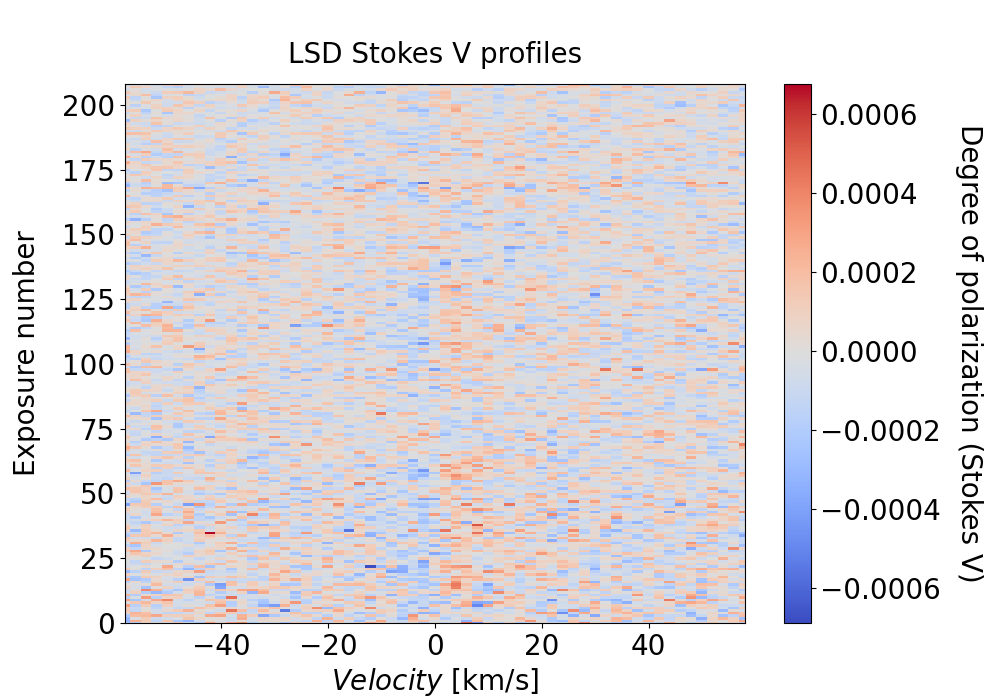}
      \caption{Same for Gl\,725B.}
    \label{fig:gl725B}
\end{figure}

\begin{figure}
   \centering
   \includegraphics[width=0.9\hsize]{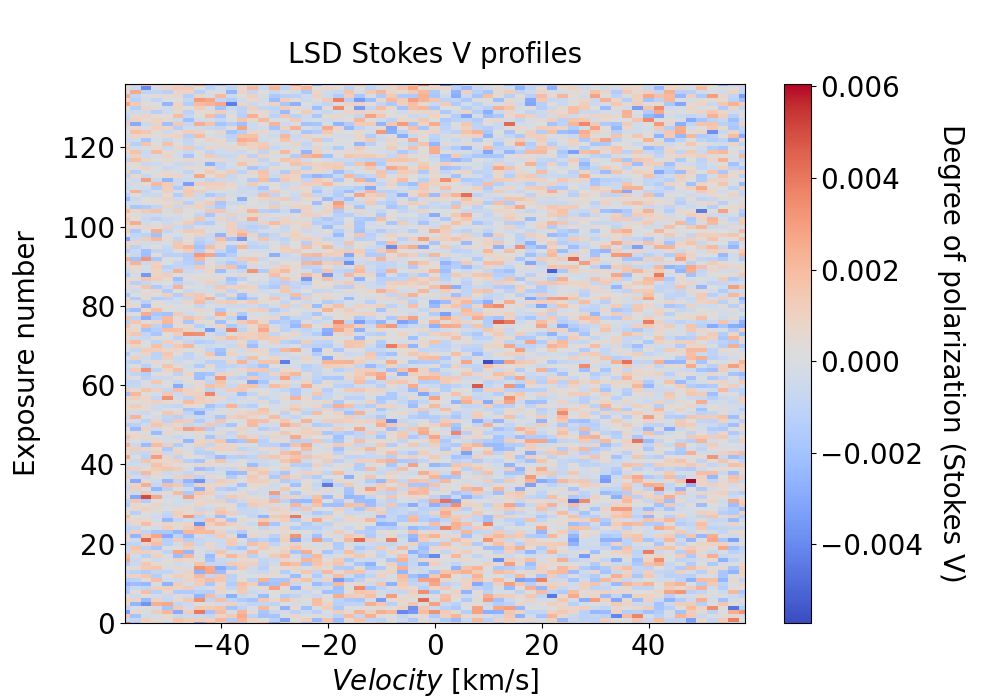}
      \caption{Same for GJ\,1105.}
    \label{fig:gl1105}
\end{figure}

\begin{figure}
   \centering
   \includegraphics[width=0.9\hsize]{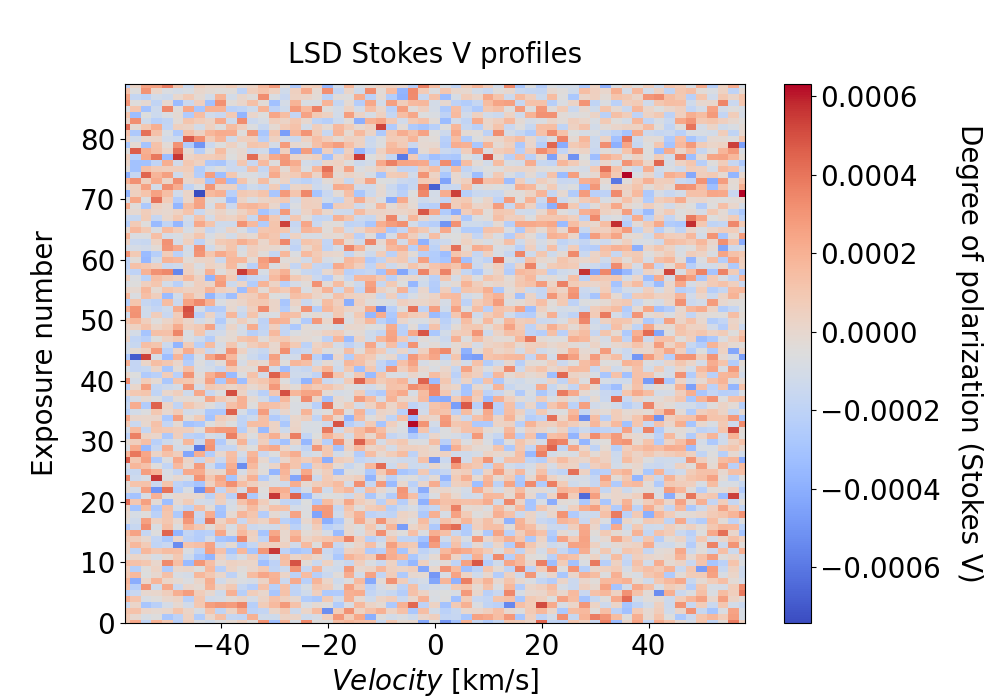}
      \caption{Same for Gl\,445.}
    \label{fig:gl445}
\end{figure}

We display here the temporal evolution of the Stokes $V$ for the 16 stars of our sample (Figs.~\ref{fig:gl338b} to \ref{fig:gl447}) for which we could not detect a periodic variation of the longitudinal magnetic field. This shows that for some stars, the Stokes V is well detected, but is constant or does not vary periodically, while for others, it is barely detected or not detected at all.

\clearpage

\begin{figure}
   \centering
   \includegraphics[width=0.9\hsize]{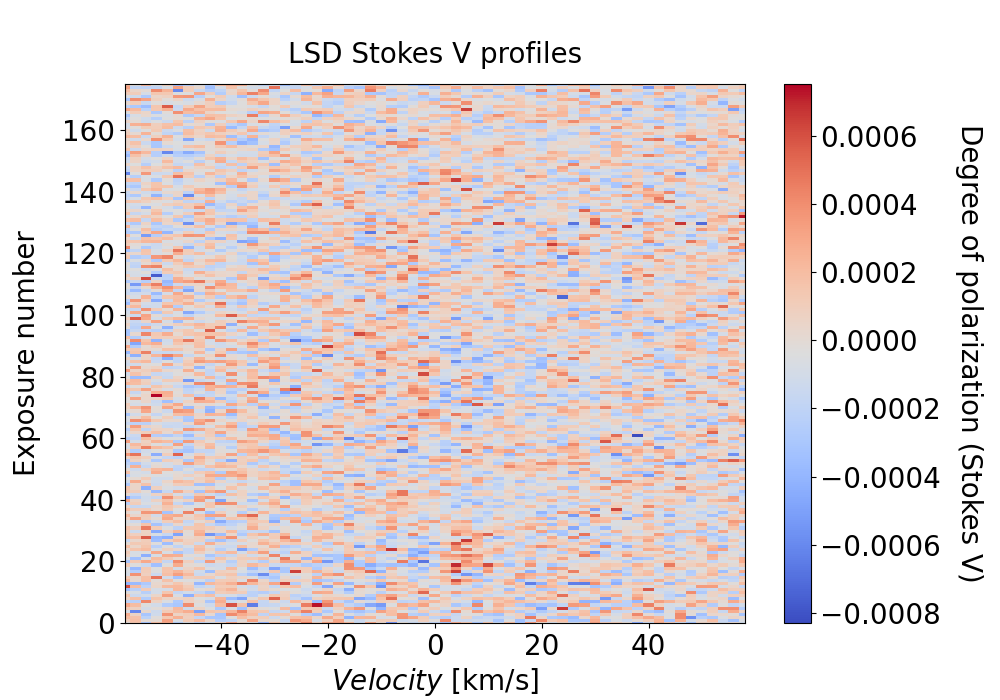}
      \caption{Same for PM\,J21463+3813.}
\end{figure}

\begin{figure}
   \centering
   \includegraphics[width=0.9\hsize]{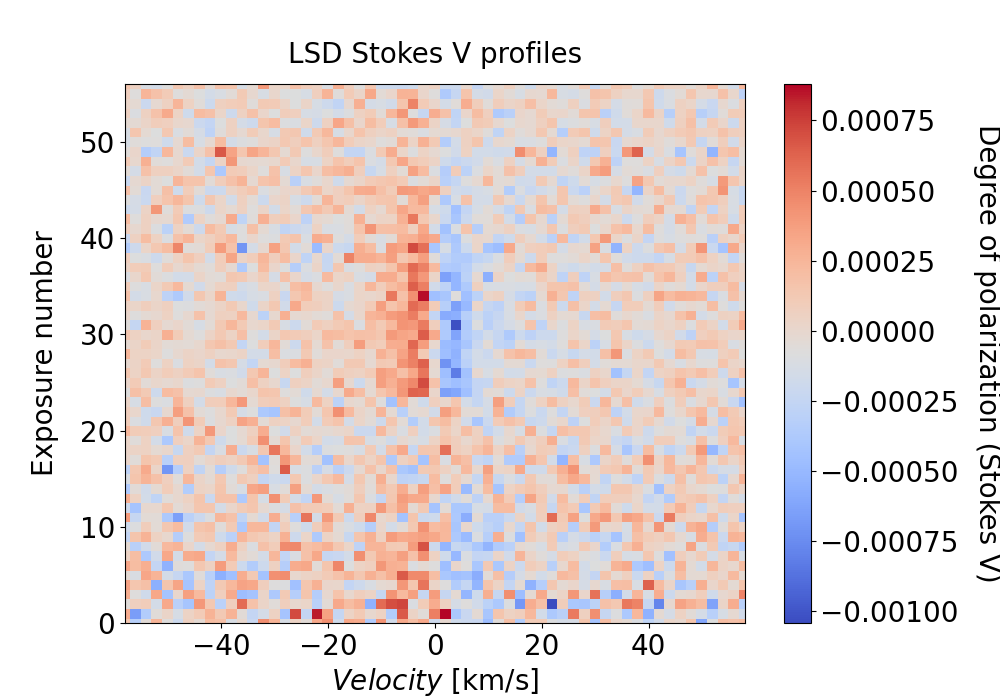}
      \caption{Same for Gl\,447.}
    \label{fig:gl447}
\end{figure}


\end{document}